\documentclass[structabstract]{aa}
\usepackage{graphicx}
\usepackage{rotating}
\usepackage{txfonts}
\usepackage{natbib}
\usepackage{lscape}
\usepackage{booktabs}
\usepackage{dcolumn}
\usepackage{float}
\bibpunct{(}{)}{;}{a}{}{,} 

\begin{document}
\renewcommand{\topfraction}{0.99}
\renewcommand{\bottomfraction}{0.99}
\renewcommand{\floatpagefraction}{0.99}
\renewcommand{\dbltopfraction}{0.99}
\renewcommand{\dblfloatpagefraction}{0.99}

\renewcommand{\topfraction}{1.99}
\renewcommand{\bottomfraction}{1.99}
\renewcommand{\floatpagefraction}{1.99}
\renewcommand{\dbltopfraction}{1.99}
\renewcommand{\dblfloatpagefraction}{1.99}
\setcounter{totalnumber}{10}
\renewcommand{\textfraction}{1.99}

\def\simless{\mathbin{\lower 3pt\hbox
{$\rlap{\raise 5pt\hbox{$\char'074$}}\mathchar"7218$}}}   
\def\simmore{\mathbin{\lower 3pt\hbox
{$\rlap{\raise 5pt\hbox{$\char'076$}}\mathchar"7218$}}}   

\title{Spectral evolution of flaring blazars from numerical simulations}
\author{C. M. Fromm\inst{1,2},  M. Perucho\inst{3,4}, P. Mimica\inst{3} and E. Ros\inst{1,3,4}}
\institute{Max-Planck-Institut f\"ur Radioastronomie, Auf dem H\"ugel 69, D-53121 Bonn, Germany\
\email{cfromm@mpifr.de}
\and Institut f\"ur Theoretische Physik, Goethe Universit\"at, Max-von-Laue-Str. 1,  D-60438 Frankfurt, Germany
\and Departament d'Astronomia i Astrof\'\i sica, Universitat de Val\`encia, Dr. Moliner 50, E-46100 Burjassot, Val\`encia, Spain
\and
Observatori Astron\`omic, Parc Cient\'{\i}fic, Universitat de Val\`encia, C/ Catedr\`atic Jos\'e Beltr\'an 2, E-46980 Paterna, Val\`encia, Spain }

\abstract
   {High resolution Very Long Baseline Interferometry (VLBI) observations of Active Galactic Nuclei (AGN) revealed traveling and stationary or quasi-stationary radio-components in several blazar jets. The traveling ones are in general interpreted as shock waves generated by pressure perturbations injected at the jet nozzle. The stationary features can be interpreted as recollimation shocks in non-pressure matched jets if they show a quasi-symmetric bump in the spectral index distribution. In some jets there may be interactions between the two kinds of shocks. These shock--shock interactions are observable with VLBI techniques, and their signature should also be imprinted on the single--dish light curves.}
   {In this paper we investigate the spectral evolution produced by the interaction between a recollimation shock with traveling shock waves to address the question of whether these interactions contribute to the observed flares and how their signature in both single--dish and VLBI observations looks like.}
   {We performed relativistic hydrodynamic (RHD) simulations of over-pressured and pressure-matched jets. To simulate the shock interaction we injected a perturbation at the jet nozzle once a steady-state was reached. We computed the non-thermal emission (including adiabatic and synchotron losses) resulting from the simulation.}
   {We show that the injection of perturbations in a jet can produce a bump in emission at GHz frequencies previous to the main flare, which is produced when the perturbation fills the jet in the observer's frame. The detailed analysis of our simulations and the non-thermal emission calculations show that interaction between a recollimation shock and traveling shock produce a typical and clear signature in both the single--dish light curves and in the VLBI observations: the flaring peaks are higher and delayed with respect to the evolution of a perturbation through a conical jet. This fact can allow to detect such interactions for stationary components lying outside of the region in where the losses are dominated by inverse Compton scattering.}
   {}

\titlerunning{Shock-shock interaction in parsec-scale jets}
\authorrunning{C. M. Fromm et al.}

\maketitle
\section{Introduction}
The kinematic analysis of high resolution VLBI images of AGN jets within long-term monitoring programs such as the MOJAVE\footnote{Monitoring of Jets in Active galactic nuclei with VLBA Experiments http://www.physics.purdue.edu/MOJAVE} program \citep{2009AJ....137.3718L} at $15\,\mathrm{GHz}$, the Boston University Blazar Monitoring\footnote{http://www.bu.edu/blazars/research.html} program \citep{2005AJ....130.1418J}  at $43\,\mathrm{GHz}$ or the TANAMI\footnote{http://pulsar.sternwarte.uni-erlangen.de/tanami/} program at 8.4~GHz and 23~GHz \citep{2010A&A...519A..45O} reveal a number of components that are stationary, i.e., constant separation from the core and nearly constant flux density. These features are typically interpreted as recollimation shocks in a over-pressured (OP, hereafter) jet and cannot be explained by traveling shock waves within a pressure-matched (PM, hereafter) and therefore, conical jet  \citep[see, e.g.,][]{1988ApJ...334..539D}.

  The signature of standing features could also be imprinted in the single-dish light curves. The spectral analysis of single-dish observations in the cm-mm and sub-millimetre regime for the blazar CTA\,102 during a major outburst leads to double hump structure in the turnover frequency -- turnover flux density plane \citep{2011A&A...531A..95F} and, at the same time, the kinematic analysis of this source from VLBI observations revealed several standing features, one of them $18\,\mathrm{pc}$ (de-projected) from the core \citep{2013A&A...551A..32F}. The spectral analysis applied to the VLBI observations exhibit an increase in the particle density and magnetic field strength at the location of the standing features \citep{2013A&A...557A.105F}. A conclusion derived from those works was that the double hump in the light curve was caused by the interaction between the traveling perturbation and a standing shock. The interaction between traveling shock waves and recollimation shocks could also be the onset of the $\gamma$-ray flares. \citet{Agudo:2010jp} and \citet{2012A&A...537A..70S} (see also references therein) combined multi-frequency observations and $43\,\mathrm{GHz}$ VLBI observations and found a correlation between the crossing of a traveling component through a stationary feature and the onset of the high energy flare. 
  
The hydrodynamics of non-pressure matched relativistic jets was studied in an analytical way by \citet{1988ApJ...334..539D}. More recently, \citet{Nalewajko:2011co} studied the formation of recollimation shocks for the case of an ultra-relativistic equation of state and a constant ambient medium density. By using the methods of characteristics, they provide several analytical solutions on the location of the pressure minimum and on the location of the first recollimation shock. \citet{1991MNRAS.250..581F} took into account a decreasing pressure in the ambient and calculated numerically the evolution of the jet. A more detailed treatment on the formation of recollimation shocks, using both, analytical approximations and numerical simulations can be found in \citet{1997MNRAS.288..833K}.
The simulations of \citet{1991MNRAS.250..581F} and \citet{1997MNRAS.288..833K} studied the formation of recollimation shocks in the context of the propagation of relativistic jets. A different approach was followed by \citet{1997ApJ...482L..33G}, who studied the propagation of relativistic shock waves in PM and OP steady-state jets and computed synthetic radio maps assuming adiabatic losses. \cite{Mimica:2009de} re-computed the emission of the simulations performed by \citet{1997ApJ...482L..33G} and included the influence of temporal and spatial radiative losses on the distribution of the relativistic particles.
So far, most of the studies focused on the propagation of the relativistic shock waves which could be connected to the observed superluminal components observed in several AGN jets. In this paper we concentrate on the interaction between traveling shock waves and recollimation shocks and the resulting spectral evolution. {With this aim, we have performed relativistic hydrodynamical numerical simulations. The current paradigm for jet launching (Blandford \& Znajek 1977) assumes that the jet is strongly magnetized close to the black hole \citep[see
e.g.,][and references therein]{2009ApJ...699.1789T,2009MNRAS.394.1182K}. The magnetization of the flow decreases further out, but it is
still possible that the flow is magnetized far away from the acceleration zone, especially in GRB case \citep{1994MNRAS.270..480T, 2001A&A...369..694S, 2003astro.ph.12347L,2006A&A...450..887G, 2011MNRAS.411.1323G,2011MNRAS.411..422L,2011IAUS..275...24L,2012MNRAS.421.2442G,2012MNRAS.422..326K}. However, since CTA102 is a blazar, \citet{2012MNRAS.421.2635M} and \citet{2014MNRAS.438.1856R} show that, to be compatible with the current blazar observations, the blazar jets are at most moderately magnetized at blazar distances ($\sigma \leq 0.01$). For these values \citet{2012MNRAS.421.2635M} show that the dynamics and emission depend on sigma only very weakly \citep[Figs. 1 and 7 in][]{2012MNRAS.421.2635M}. Therefore the assumption about the non-magnetized jet dynamics with $\epsilon_B\sim0.1$ at those distances is justified . Further out of that zone, at distances of interest in this paper, the magnetization should be even smaller.}

The organization of this work is the following: In Sect.~\ref{setup} we introduce our numerical setup. The results of the simulations and the non-thermal emission calculations are presented in Sect.~\ref{hydro} and in Sect.~\ref{emission}. The discussion of our results is provided in Sect.~\ref{disc}. Throughout the paper we use an ideal equation of state $p=(\hat{\gamma}-1)\epsilon\rho$, with pressure, $p$, adiabatic index, $\hat{\gamma}$, specific internal energy, $\epsilon$, and density, $\rho$.

\section{RHD Simulations}
\label{setup}
We performed several 2D axisymmetric simulations of supersonic relativistic hydrodynamical jets using the finite-difference code \textit{Ratpenat} \citep[for more details see][and references therein]{Perucho:2010ht}. The simulations were performed on up to 64 processors at the local cluster at the Max-Planck-Institute for Radio Astronomy (MPIfR) and at \textit{Tirant}, the Valencian Node of the Spanish Supercomputing Network (RES). 

\subsection{Simulation set-up}
\begin{figure}[h!]
\resizebox{\hsize}{!}{\includegraphics{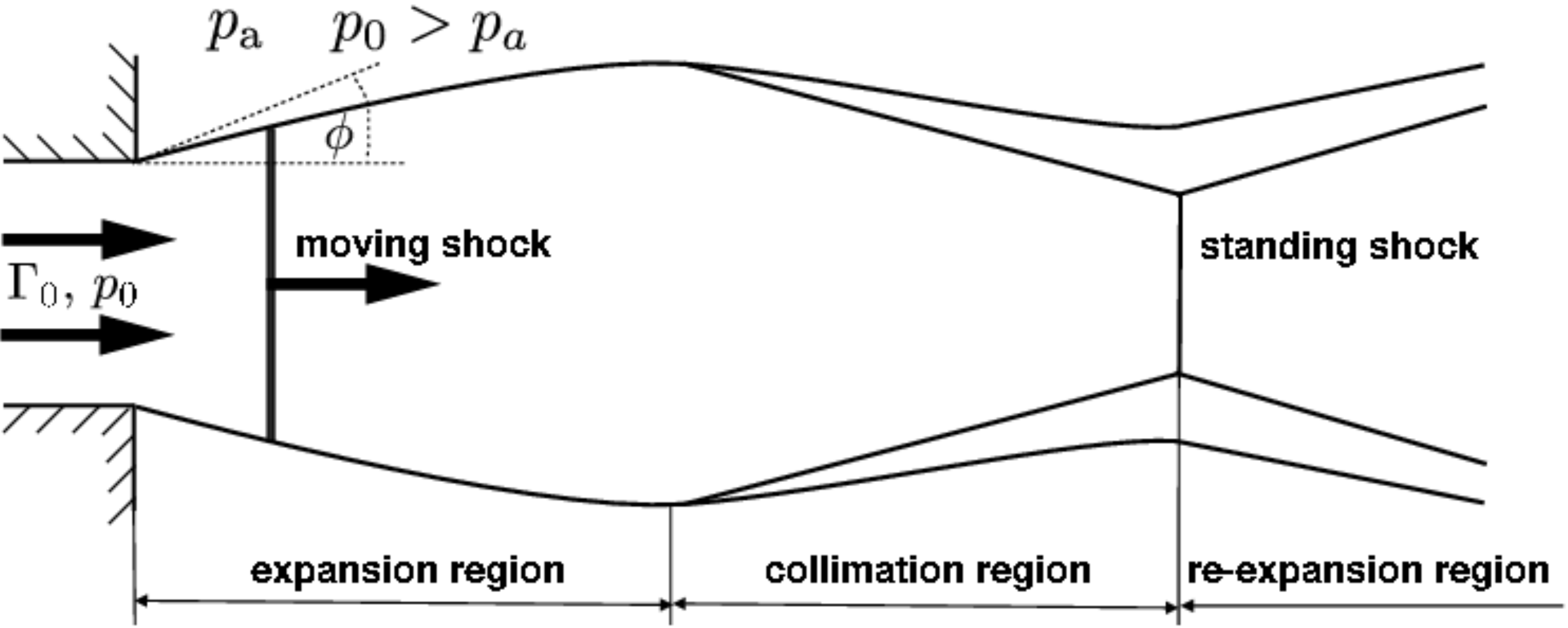}} 
\caption{Sketch of an OP jet with characteristic parameters and regions (adopted from \citealt{1988ApJ...334..539D}).}
\label{sketch} 
\end{figure}

The numerical grid includes 320 cells in the radial direction and 9600 cells in the axial direction. Using a numerical resolution of 32 cells per jet radius ($R_j$), the grid covers $10\, R_j \times 300\,R_j$. We define the $z$-axis in direction of the jet propagation and the $x$-axis as the radial axis in cylindrical coordinates. The boundary conditions are reflection at the jet axis, injection at the jet nozzle and outflow conditions elsewhere. The basic setup of our simulation for an OP jet is sketched in Fig.~\ref{sketch}. The initial parameters at the jet nozzle are the velocity of the jet, $v_\mathrm{b}$, the bulk Lorentz factor, $\Gamma$, the classical Mach number of the jet, $M$, the density of the jet, $\rho_{b}$, the adiabatic index, $\gamma$, and the initial pressure mismatch between the jet and the ambient medium, $d_k=p_{b}/p_{a}$. The pressure, $p_{b}$ at the jet nozzle is computed from the Mach number using an ideal-gas equation of state. Since we are mainly interested in the first traveling shock--recollimation shock interaction we used a homogeneous ambient medium \citep[see, e.g.,][]{2015sebc.book.....F}. In order to study shock-jet interaction from a single traveling shock we additionally simulate a PM jet, $d_k=1$, in a decreasing pressure ambient, which leads to the formation of a conical jet without recollimation shocks. We model the decrease in the ambient medium pressure using a pressure profile presented in \citet{1997ApJ...482L..33G}:
\begin{equation}
p_a(z)=\frac{p_b}{d_k}\left[1+\left(\frac{z}{z_c}\right)^n\right]^{\frac{m}{n}},
\label{pamb}
\end{equation}
where $z_c$ can be considered as the ''spatial scale" and the exponents $n$ and $m$ control the steepening of the ambient pressure.
The initial conditions, given in code units (speed of light $c=1$, jet radius $R_j$, and ambient medium density, $\rho_a=1$) for both simulations are listed in Table~\ref{para}.

\begin{table}[h!]
\caption{Initial parameters for the simulations in code units}
\label{para}
\centering
\begin{tabular}{c c c c c c c c c c}
\hline\hline
R$_{b}$ &	$v_{b}$ 	&	$d_k$	&	$\Gamma$ 	&	$\rho_b$		&	M	& $\hat\gamma$ & $z_c$ & $m$ & $n$\\ 
$[1]$	&	[c]	&[1]		&	[1]		&	[$\rho_a$] &	[1]		&	[1]		  	& [$R_j$] & [1] & [1] \\
\hline
1		&  0.99652	&	3		&	12	&	0.02&	3.0		&$13/9$	& 0	& 0 &0\\
1		&  0.99652	&	1		&	12	&	0.02	&	3.0		&$13/9$	& 50 	& 1 & 2\\
\hline
\end{tabular}
\end{table}

Once the steady state is reached (after approximately 5 longitudinal grid crossing times), we injected a perturbation at the jet nozzle. In order to chase a shock wave to develop, we increased the pressure and density of the perturbation  by a factor of  4 (compared to the steady state pressure and density), while keeping the same velocity as the jet flow. The parameters for the perturbation, in code units, are presented in Table~\ref{paras}. 

\begin{table}[h!]
\caption{Perturbation parameters for the simulations in code units}
\label{paras}
\centering
\begin{tabular}{c c c c }
\hline\hline
 $\Delta t$	&	$v_{p}$ 	&	$\rho_p$	 & $p_p$\\ 
 $[R_j/c]$	&	[c]	&	[$\rho_a$] &	[$\rho_a c^2$] \\
\hline
0.2		&  0.99652	&	0.08 &	0.008	\\
\hline
\end{tabular}
\end{table}

\section{Results}
\label{results}
\subsection{RHD}
\label{hydro}

\begin{figure}[t!]
\includegraphics[width=\columnwidth]{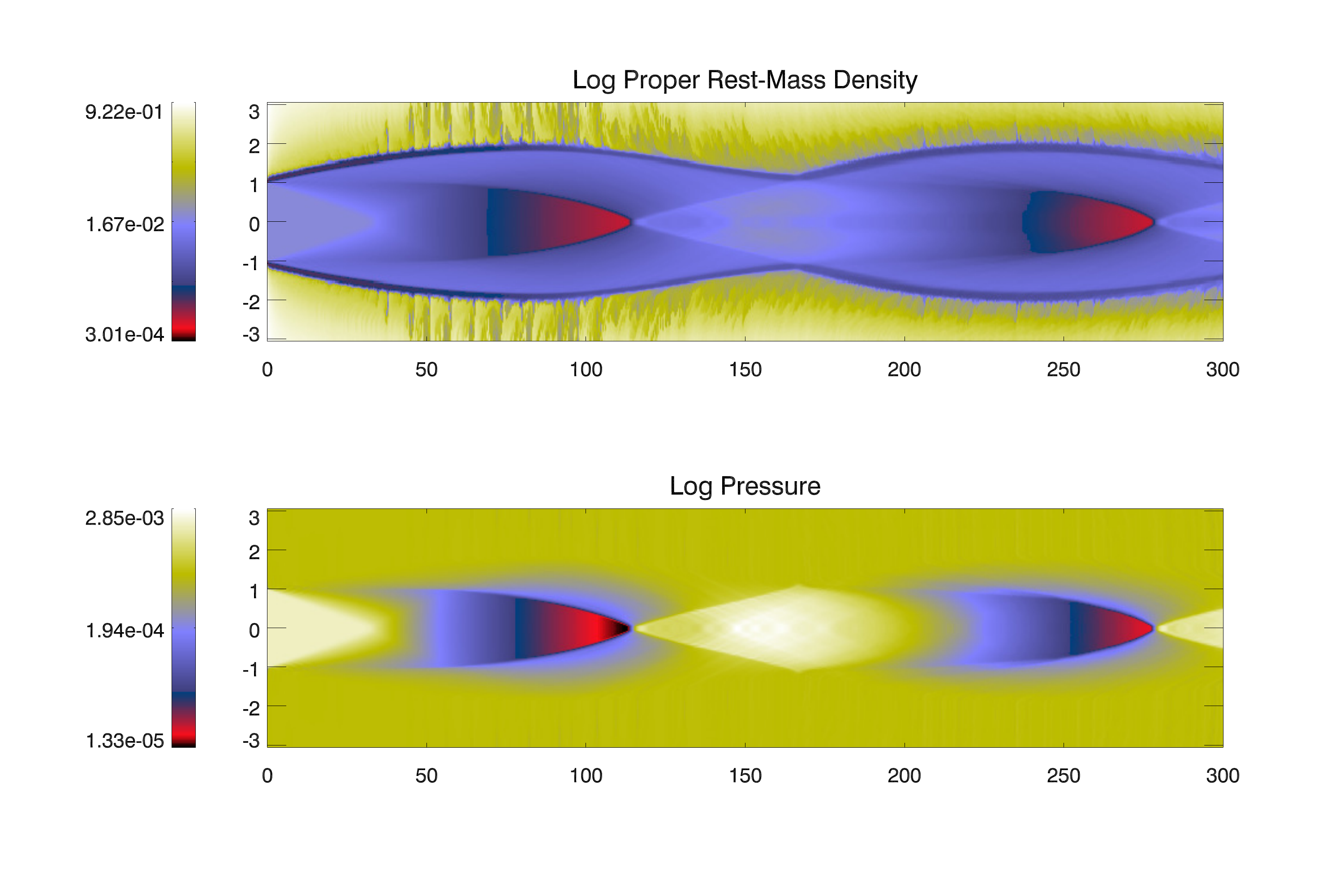} 
\caption{Steady-state results for the simulation of the OP jet. The top panel shows the 2D distribution of the logarithm of the rest mass density in units of $\rho_a$ and the bottom panel the logarithm of the pressure in units of $\rho_a c^2$}
\label{dk3steady} 
\end{figure}

\begin{figure}[t!]
\includegraphics[width=\columnwidth]{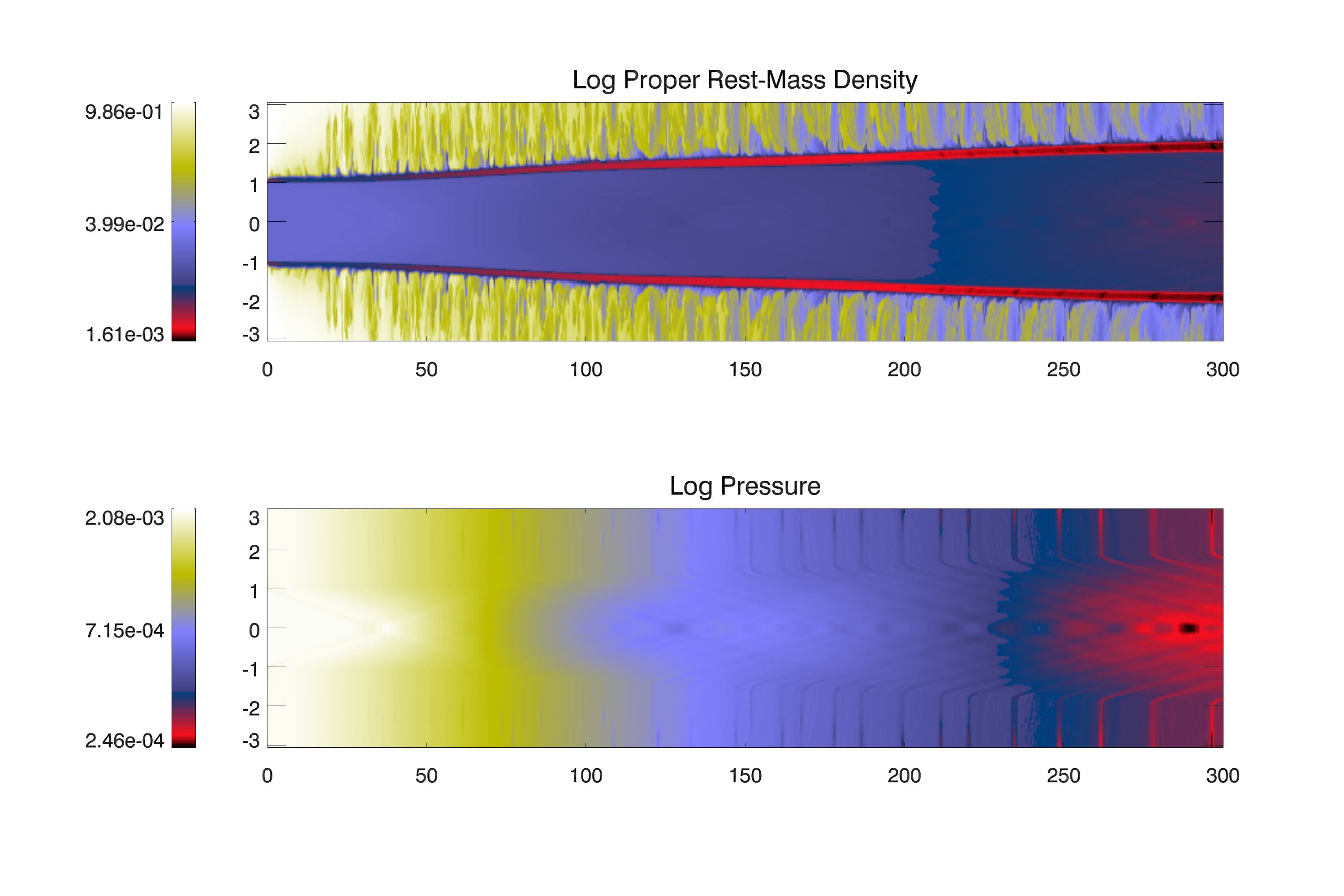} 
\caption{Steady-state results for the simulation of a PM jet. The top panel shows the 2D distribution of the logarithm of the rest mass density in units of $\rho_a$ and the bottom panel the logarithm of the pressure in units of $\rho_a c^2$}
\label{dk1steady} 
\end{figure}

Figure~\ref{dk3steady} shows the 2D distribution of rest mass density (top) and pressure (bottom) in case of the OP jet ($d_k=3$) for the steady state. Due to the pressure mismatch at the jet nozzle, two shock waves form at the discontinuity between the jet and the ambient. One of them propagates outwards in the radial direction and the other propagates towards the axis. Between them, a rarefaction region forms in which the flow expands radially until pressure equilibrium between the the jet and the ambient medium is established. This state is first reached at the jet boundary, and leads to the formation of an inward traveling sound wave. Due to the finite speed of the waves, the inner layers of the jet will continue expanding while the outer ones are already being collimated.  This expansion of each inner layer stops as soon as the waves cross it. The recollimation shock, related to the shock wave that propagates towards the axis, occurs at different locations for different values of the radial coordinate of the stream line: The expansion and recollimation of the flow is clearly visible in Fig.~\ref{dk3steady}. The recollimation shock reaches the axis at $z=110\,R_j$. At this position there is a local maximum in pressure and density and the flow emerging from this region expands again due to increased pressure. In other words, the recollimation shock can be considered as a new ``jet nozzle'', and the process begins anew. In our case the second recollimation shock is formed at $z=270\,R_j$.  A different scenario is obtained for the PM jet ($d_k=1$). The distribution of the rest mass density and the pressure is smooth along the jet (see Fig.~\ref{dk1steady}), as the jet expands, adapting to the ambient pressure (compare Fig.~\ref{dk1steady} to Fig.~\ref{dk3steady}).

\begin{figure}[t!]
\includegraphics[width=\columnwidth]{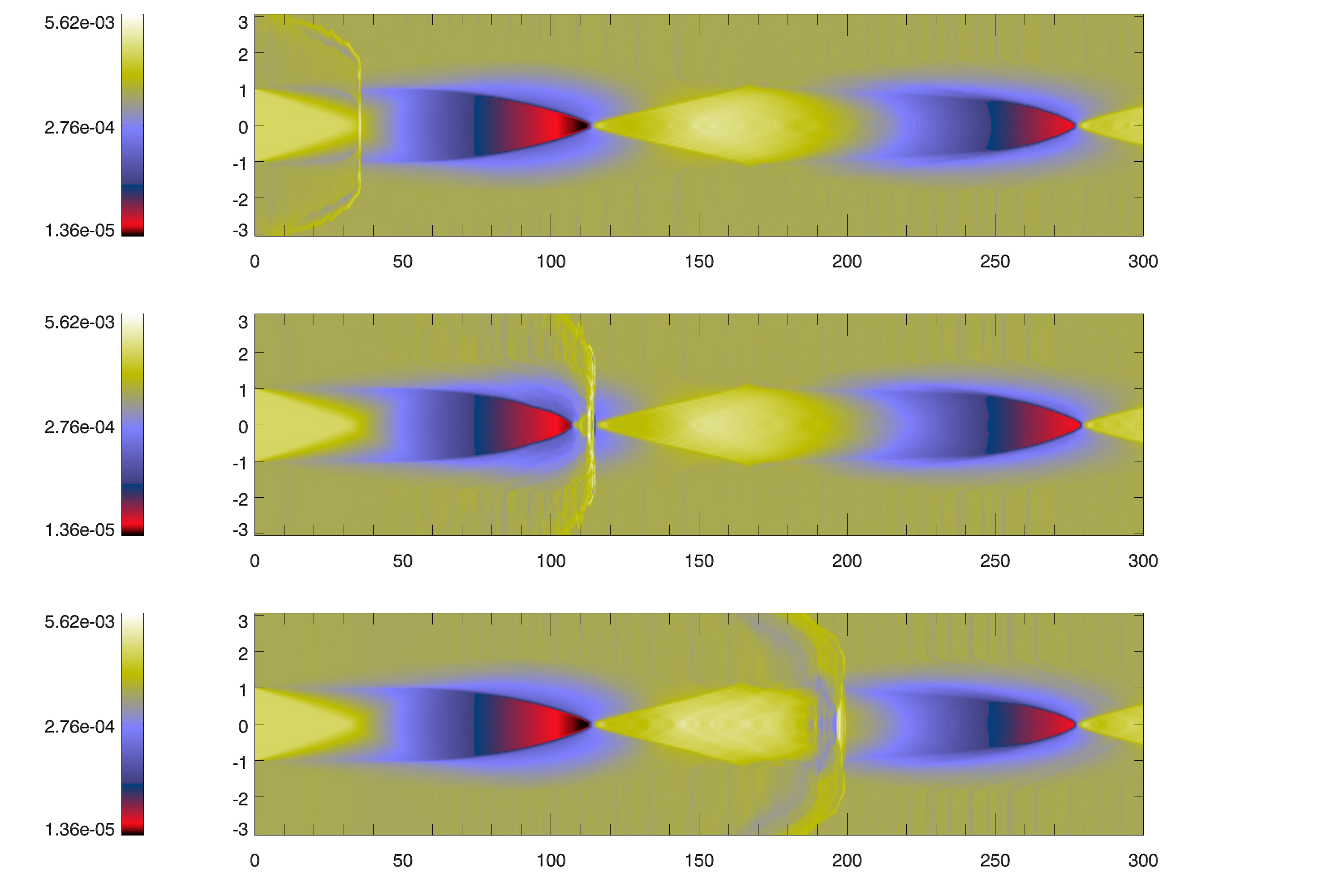} 
\caption{Snapshots for the propagation of a perturbation in an OP jet. The panels shows the 2D distribution of the logarithm of the pressure in units of $\rho_ac^2$.} 
\label{dk3ss} 
\end{figure}

\begin{figure}[t!]
\includegraphics[width=\columnwidth]{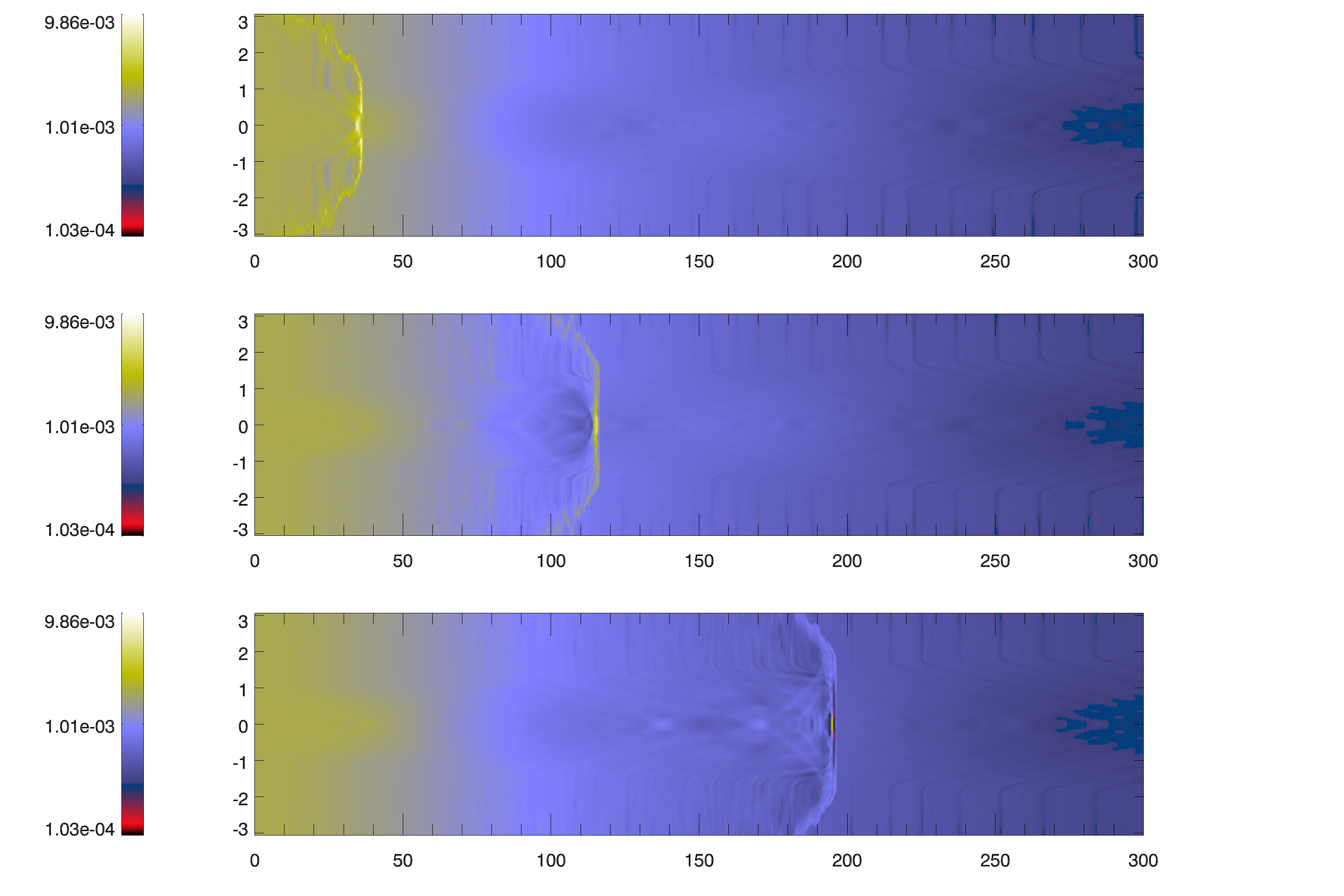} 
\caption{Snapshots for the propagation of a perturbation in a PM jet. The panels shows the 2D distribution of the logarithm of the pressure in units of $\rho_ac^2$}
\label{dk1ss} 
\end{figure}

\begin{figure}[t!]
\includegraphics[width=\columnwidth]{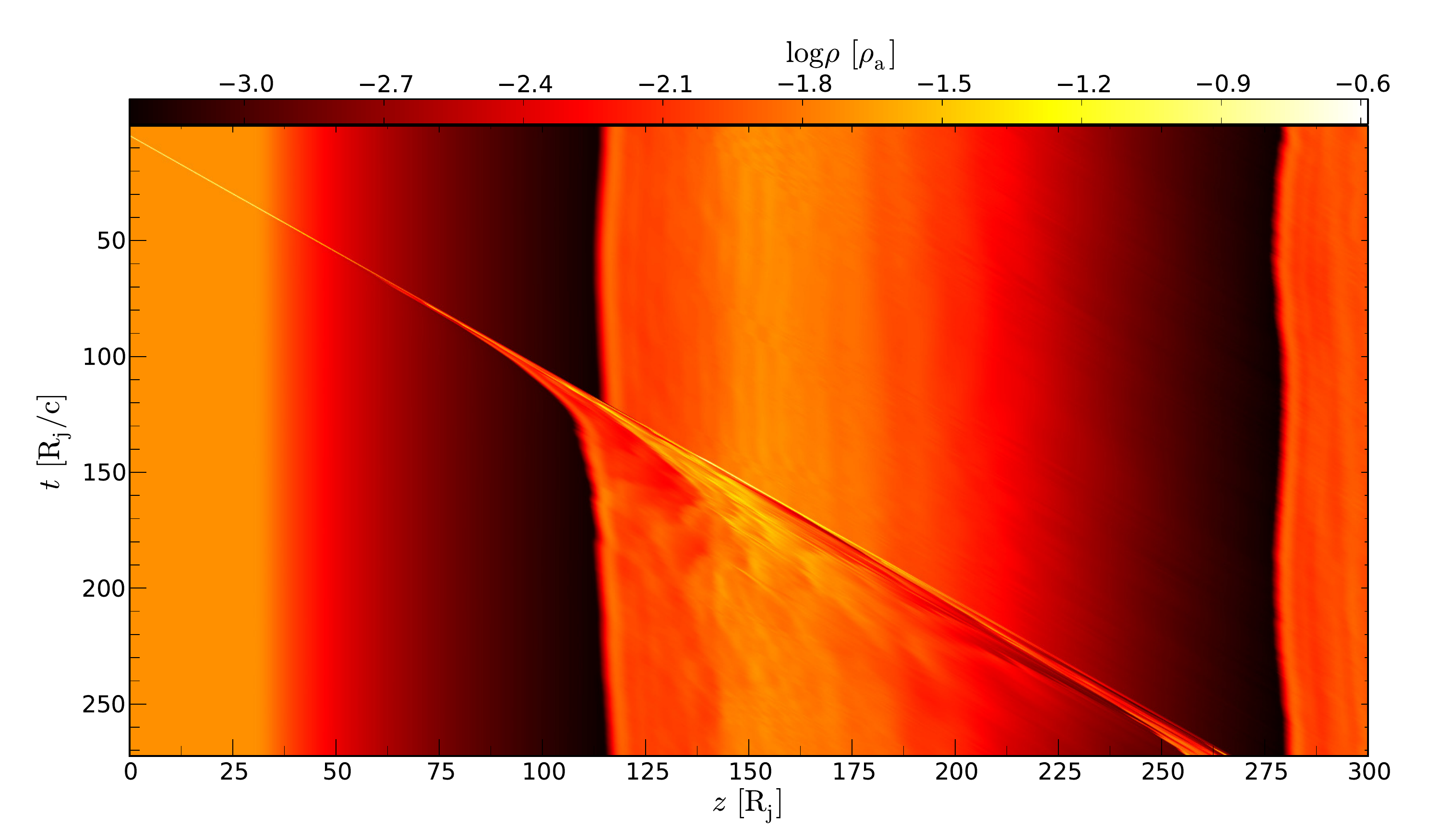} 
\caption{Time-Space plot for the variation of the axial density for the OP jet.}
\label{tdk3ss} 
\end{figure}

\begin{figure}[t!]
\includegraphics[width=\columnwidth]{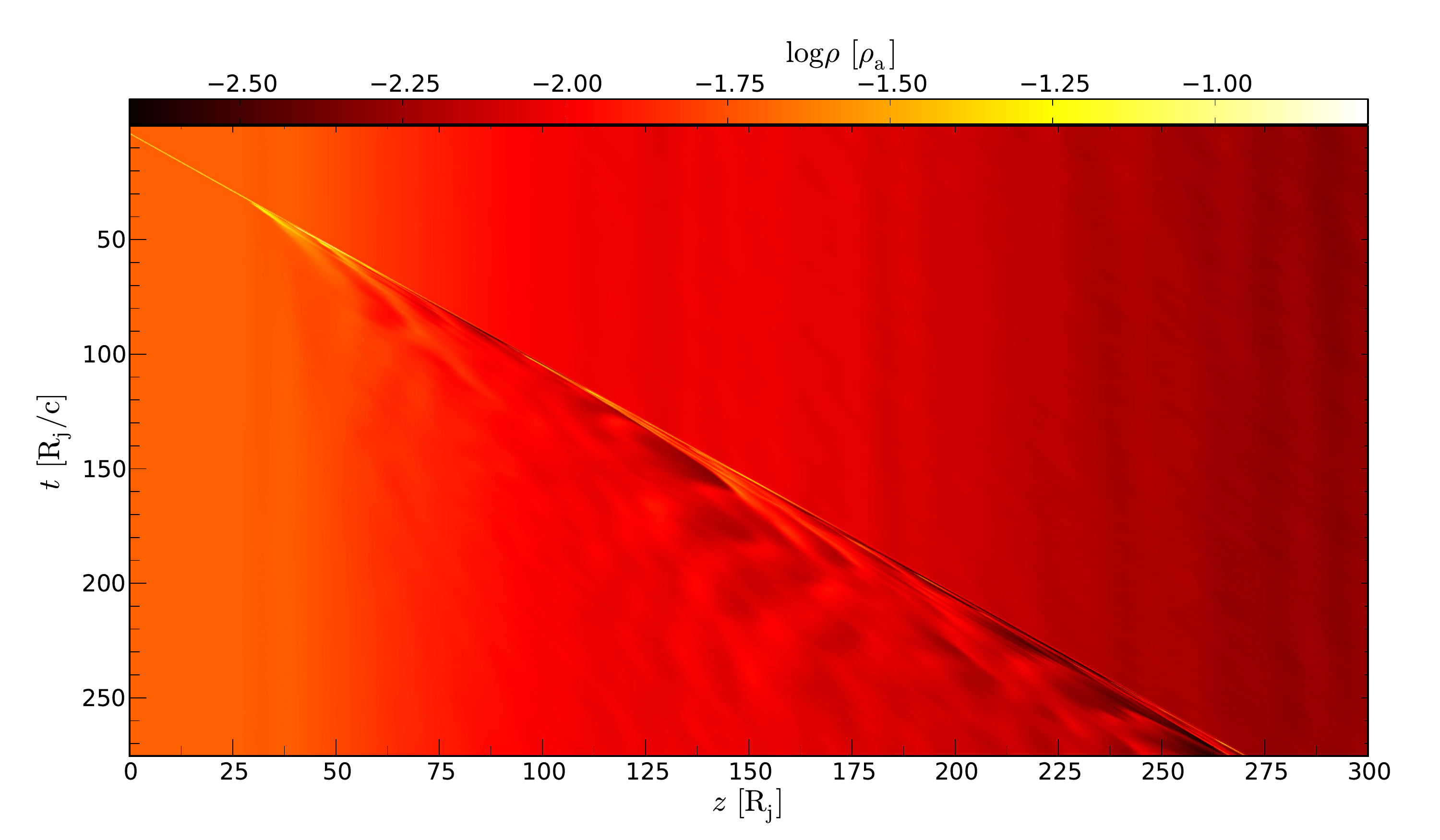} 
\caption{Time-Space plot for the variation of the axial density for the PM jet.}
\label{tdk1ss} 
\end{figure}

Once the perturbation is injected, a shock wave (forward) and a rarefaction wave (reverse) are generated \citep[see, e.g.,][]{1999LRR.....2....3M}. The jet material swept up by the shock wave is compressed (pressure and density increase), while the crossing of the rarefaction wave induces a decrease in both quantities. As an example of the propagation of a perturbation, Fig. \ref{dk3ss} (OP jet) and Fig.~\ref{dk1ss} (PM jet) show the variation in pressure at three selected times. The entire evolution of the axial density during the propagation of the shock wave is presented in Figs.~\ref{tdk3ss} and \ref{tdk1ss}. The variation in the pressure and the rest mass density during the propagation of the shock during the first $50\,\mathrm{R_j}$ is similar in the OP and the PM jet until $50\,\mathrm{R_j}$. The opening of the jet leads to an expansion of the shock wave and the compression of the gas produced by the shock falls with the distance. While the PM jet continues expanding (adapting to the decreasing ambient medium pressure), the OP jet starts collimating and forms a strong re-collimation shock. The differences in the properties of the underlying jet change the evolution of the perturbation significantly. The compression induced by shock wave continues to decrease in the PM jet (see Fig. \ref{tdk1ss}). In contrast to this, in the OP jet the increase in the compression of the pressure and density during and after the interaction between the shock wave and the recollimation shock is seen at $z\approx120\,\mathrm{R_j}$ (see Fig. \ref{tdk3ss}). In addition to the differences in the compression of the underlying flow, the trailing features (secondary perturbations generated in the wake of the main one) are stronger and broader in the OP jet than in the PM jet (best seen at $t\approx150\,\mathrm{R_j/c}$ in Figs.~\ref{tdk3ss} and \ref{tdk1ss}), and appear associated to the interaction between the perturbation and the standing shock in the former. This is the physical setup and in the next section we proceed to compute the non-thermal emission from the simulated jets.

\subsection{Emission}
\label{emission}
\subsubsection{SPEV setup}
We used the code SPEV \citep{Mimica:2009de} to compute the radio
emission. SPEV injects a representative population of Lagrangian
particles (LPs) at the jet nozzle and evolves them in space and time
according to the conditions of the underlying relativistic
outflow. SPEV takes into account radiative losses and adiabatic
effects of the emitting non-thermal electrons (NTEs), and computes their time- and
frequency-dependent synchrotron emission. It then performs the
radiative transfer taking into account special relativistic effects
and time delays. The radiation is collected on a virtual detector
consisting of a number of pixels.

{Regarding the possible effect of the implicit assumption in our simulations of a purely hydrodynamical flow, the absence of a dynamically relevant magnetic field could have an important effect in the emission calculations. In particular, it has been shown by Sironi et al. (2015) that strongly magnetized flows are not very efficient in accelerating non-thermal electrons to high energies. However, we have argued at the introduction that the jets are probably only weakly magnetized at the scales that we study, and, in addition, we also try to take this effect into account, even if small in our case, by assuming a finite maximum electron energy which is inversely proportional to the square root of the magnetic field strength (see Eq. \ref{gammamax}):

\begin{equation}
\label{gammamax}
 \gamma_{max} = \left(\frac{3m_e^2 c^4}{4\pi a_{\rm acc} e^3
B}\right)^{1/2}\,,
\end{equation}

where $\gamma_\mathrm{max}$ is the upper cut-off of the injected non-thermal energy distribution, $m_e$ and $c$ are the electron mass and the speed of light, $B$ is the comoving magnetic field, and $a_{\rm acc}$ is the acceleration efficiency parameter \citep[see][]{2010ApJ...711..445B}. We use $a_{\rm acc}= 10^6$ \citep[similar to]{MimicaAloy:2012}. Furthermore, we assume that the emitting particles are accelerated in the inner parts of the jet (not simulated by us), and are injected though the nozzle at the inner boundary in our simulation.}

\begin{figure}[h!]
\includegraphics[width=\columnwidth]{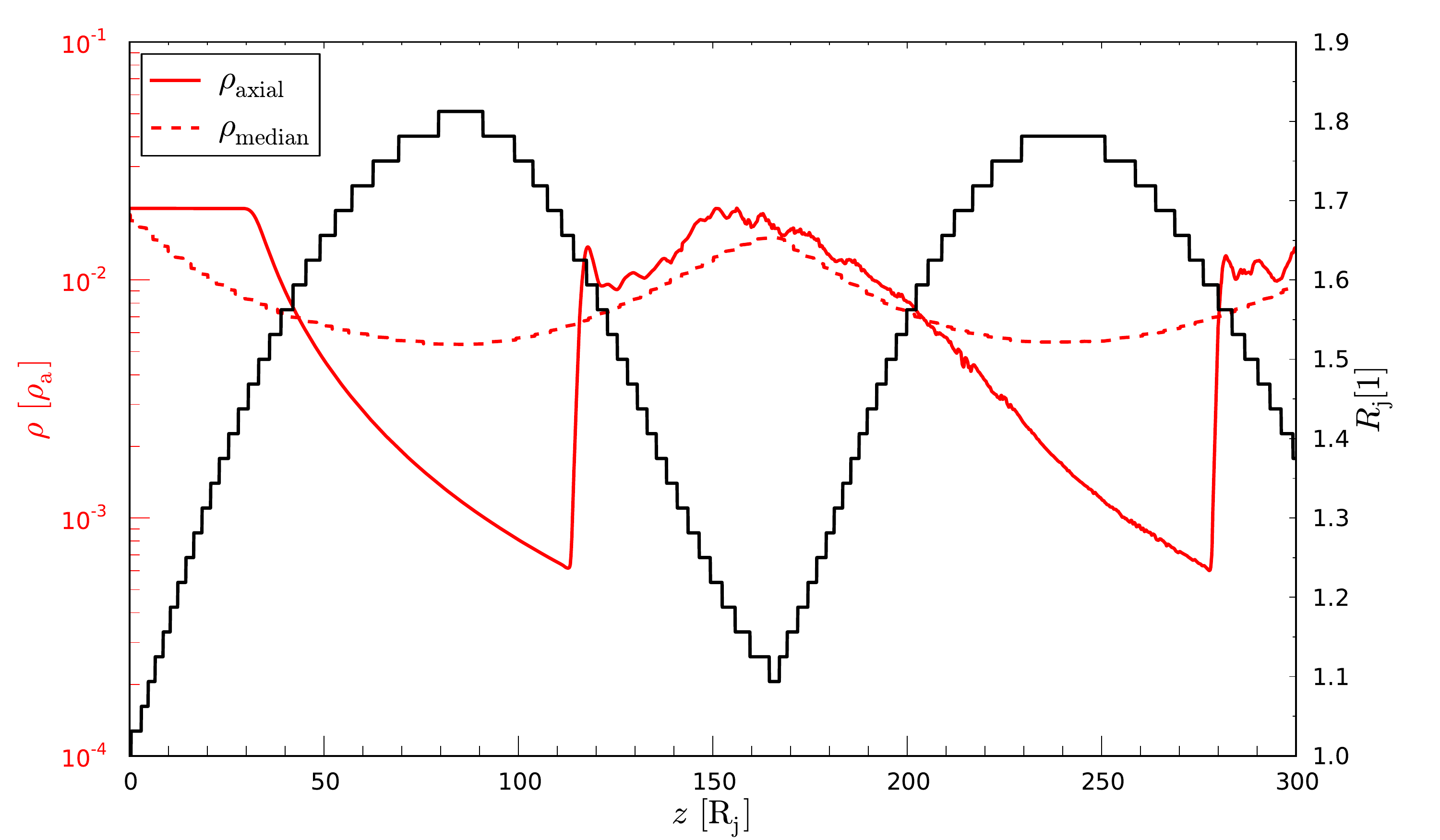} 
\caption{Variation of the axial, $\rho_\mathrm{axial}$, and the cross-section averaged rest mass density, $\rho_\mathrm{median}$ along the jet (red lines) and the variation of the jet radius along the jet (black line) for the OP jet.}
\label{averaged} 
\end{figure}

In this work we use $\simeq 3000$ snapshots produced by Ratpenat as a
dynamic background for the evolution of LPs. In order to make
the calculations feasible, we averaged the hydrodynamical quantities in
the direction transverse to the jet, but kept track of the jet
radius at each point. This allows us to reduce the amount of intermediate data we need to
store: at each point along the jet it is only necessary to store one set of
hydrodynamical quantities (density, pressure, and velocity). However, since
the radius is recorded as well, we can reconstruct the jet geometry at any  point and time (see Fig. \ref{averaged}).

The data are used by SPEV to continuously inject $48$ LP families at the jet nozzle \citep[see Sec. 3.3 of][for a detailed explanation of this process]{Mimica:2009de}. The LP families are distributed transversally along the jet nozzle and their trajectories follow the jet geometry (i.e., where the jet gets narrower the trajectories come closer together, and vice-versa). We assume that each LP represents a volume uniformly filled with NTEs, an isotropic NTE velocity distribution within each LP and only track the NTE energy distribution. With this in mind, we initially we assume a power-law distribution:
\[ n(\gamma) = n_0 (\gamma/\gamma_{\rm min})^{-s};\ \gamma_{\rm
 min}\leq\gamma\leq\gamma_{\rm max} \]where $n(\gamma)$ is the differential number of NTEs with Lorentz
factor $\gamma$, $n_0$ and $s$ are the normalization and the power-law
index and $\gamma_{\rm min}$ and $\gamma_{\rm max}$ are the lower and
upper cutoffs of the distribution. We assume that the total number
density of injected NTEs is proportional to the fluid number
density, where $\zeta_e$ is the ratio of non-thermal particles to thermal particles \citep[see e.g.,][]{Boettcher:2010,MimicaAloy:2012},
\[ \int_{\gamma_{\rm min}}^{\gamma_{\rm max}}\ {\mathrm{d}}\gamma\
n(\gamma) = \zeta_e \rho / m_p \ , \]
and that the total NTE energy density is
proportional to the fluid internal energy density $\varepsilon$,
\[ \int_{\gamma_{\rm min}}^{\gamma_{\rm max}}\ {\mathrm{d}}\gamma\
n(\gamma) \gamma m_e c^2 = \epsilon_e \varepsilon \, , \] where $\epsilon_e$ is ratio between the energy stored the thermal particles and in the non-thermal ones.
The upper cutoff of the spectral distribution is determined by the balance between synchrotron cooling and
acceleration timescales \citep[see e.g.,][]{Mimica:2010,MimicaAloy:2012},
\[ \gamma_{\rm max} = (3 m_e^2 c^4 / 4\pi e^3 B)^{1/2}\, , \] 
where $B$ is the comoving magnetic field (see below). Combining the equations above and inserting for $\varepsilon$ the equation of state (here ideal equation of state), leads to the following relation for the lower electron Lorentz factor, $\gamma_{\rm min}$:

\[ \gamma_\mathrm{min}=\left\{ \begin{array}{ll} 
\frac{p}{\rho} \frac{m_p}{m_ec^2}\frac{(s-2)}{(s-1)(\hat{\gamma}-1)}\frac{\epsilon_e}{\zeta_e} & \textrm{if }s>2\\
\left[\frac{p}{\rho} \frac{m_p}{m_ec^2}\frac{(2-s)}{(s-1)(\hat{\gamma}-1)}\frac{\epsilon_e}{\zeta_e} \gamma_\mathrm{max}^{s-2}\right]^{1/(s-1)} & \textrm{if }1<s<2\\ 
\frac{p}{\rho}\frac{\epsilon_e}{\zeta_e} \frac{m_p}{m_ec^2(\hat{\gamma}-1)}/\ln\left(\frac{\gamma_\mathrm{max}}{\gamma_\mathrm{min}}\right) & \textrm{if } s=2
\end{array}\right. \]

The preceding three equations, together with the assumption that $s$ is constant, allow us to determine the four parameters needed to compute the injected NTE spectrum\footnote{We note that, depending on the fluid properties at the injection point, it might happen that $\gamma_{\rm min} \leq 1$. In this case we follow the prescriptions of Sec.~2.1.1 of \citet{Sironi:2013} for the ``deep Newtonian'' regime: we set $\gamma_{\rm min}=1$ and recompute $n_0$.}. The magnetic field is assumed to be tangled, and its energy density is assumed to be be proportional to the fluid internal energy density,
\[ B = (8\pi \epsilon_B \varepsilon)^{1/2}\ , \]with $\epsilon_B$ as the equipartition ratio between the magnetic field energy density and the internal energy density of the thermal particles.
The injected LPs gradually fill the jet volume so that at the end
of the calculation the total number of LPs which emit is $\simeq
4\times 10^8$. The spatial resolution of the virtual detector is $1.5 \times
10^{16}$ cm / pixel ($\simeq 0.005$ pc / pixel). We computed the images
for $200$ temporal snapshots spanning $4$ years of the observer
time. The frequencies at which we computed the images are the standard VLBI bands and parameters used for the calculation of the emission are presented in Table \ref{paras}.

\begin{table}[h!]
\caption{Parameters for the emission simulations}
\label{paras}
\centering
\begin{tabular}{c c c c c c}
\hline\hline
 $\rho_a $	&	$R_j$ 	&	$\epsilon_B$	 & $\epsilon_e$ & $\zeta_e$ &z\\ 
 $[\mathrm{g/cm^3}]$	&	[cm]	&	[1] &	[1]  & [1]  &[1]\\
\hline
$1.67\times 10^{-21}$		&  $3.08\times 10^{17}$	&	0.1 &	0.5	& 1.0 &1.0\\
\hline
\end{tabular}
\end{table}

\subsubsection{Single dish light curves and light curve parameters}
\label{sec:single-dish}
\begin{figure}[t!]
\includegraphics[width=\columnwidth]{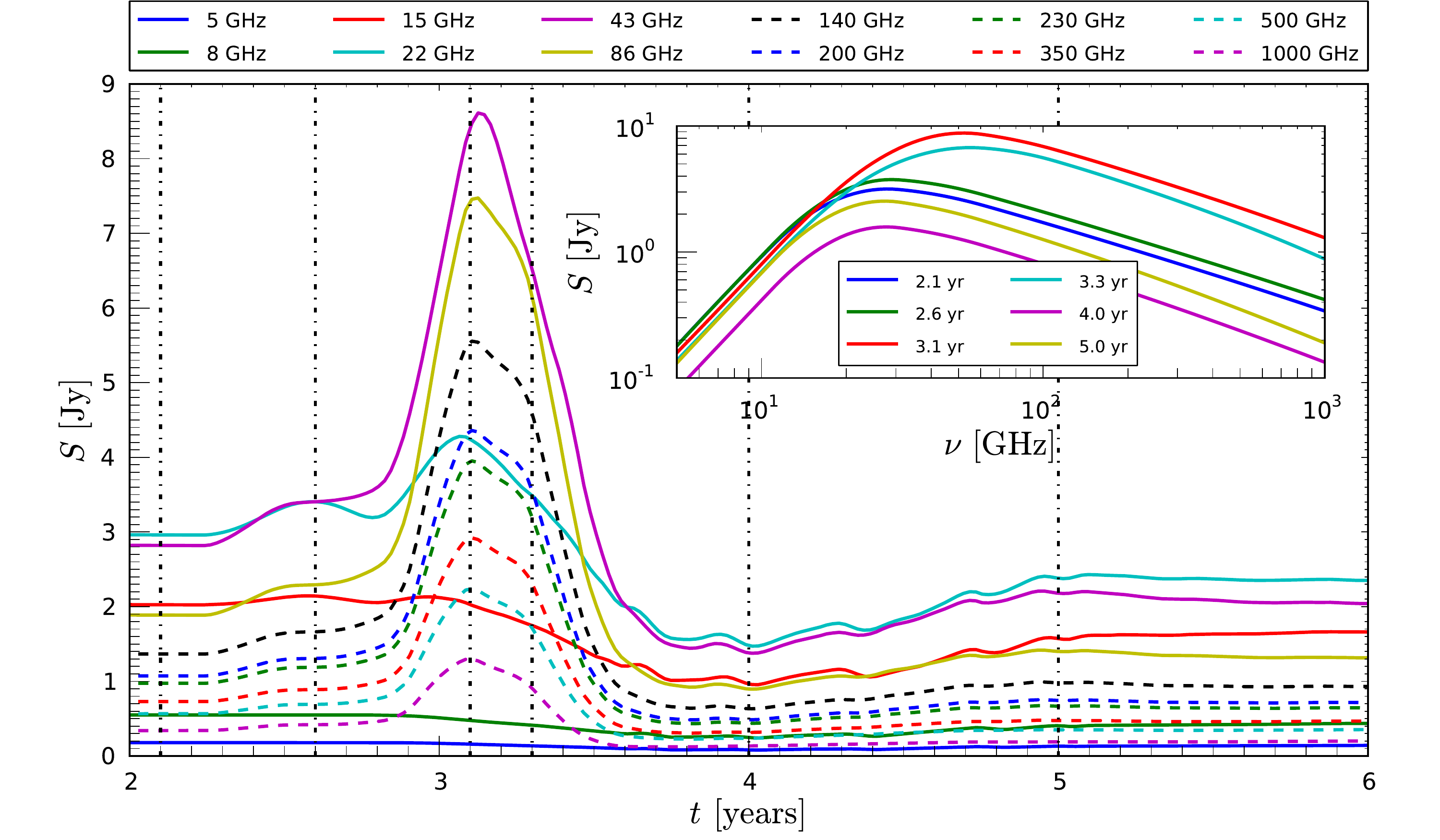} 
\caption{Single-dish light curves for several frequencies computed for the OP jet. The time is given in the observers frame. Notice that the first flux density variations are observed two years after the injection of the perturbation (second dashed vertical line from the left). The inlet shows the variation of the spectrum for six different times indicated by dashed-dotted vertical lines. }
\label{lcdk3} 
\end{figure}

\begin{figure}[t!]
\includegraphics[width=\columnwidth]{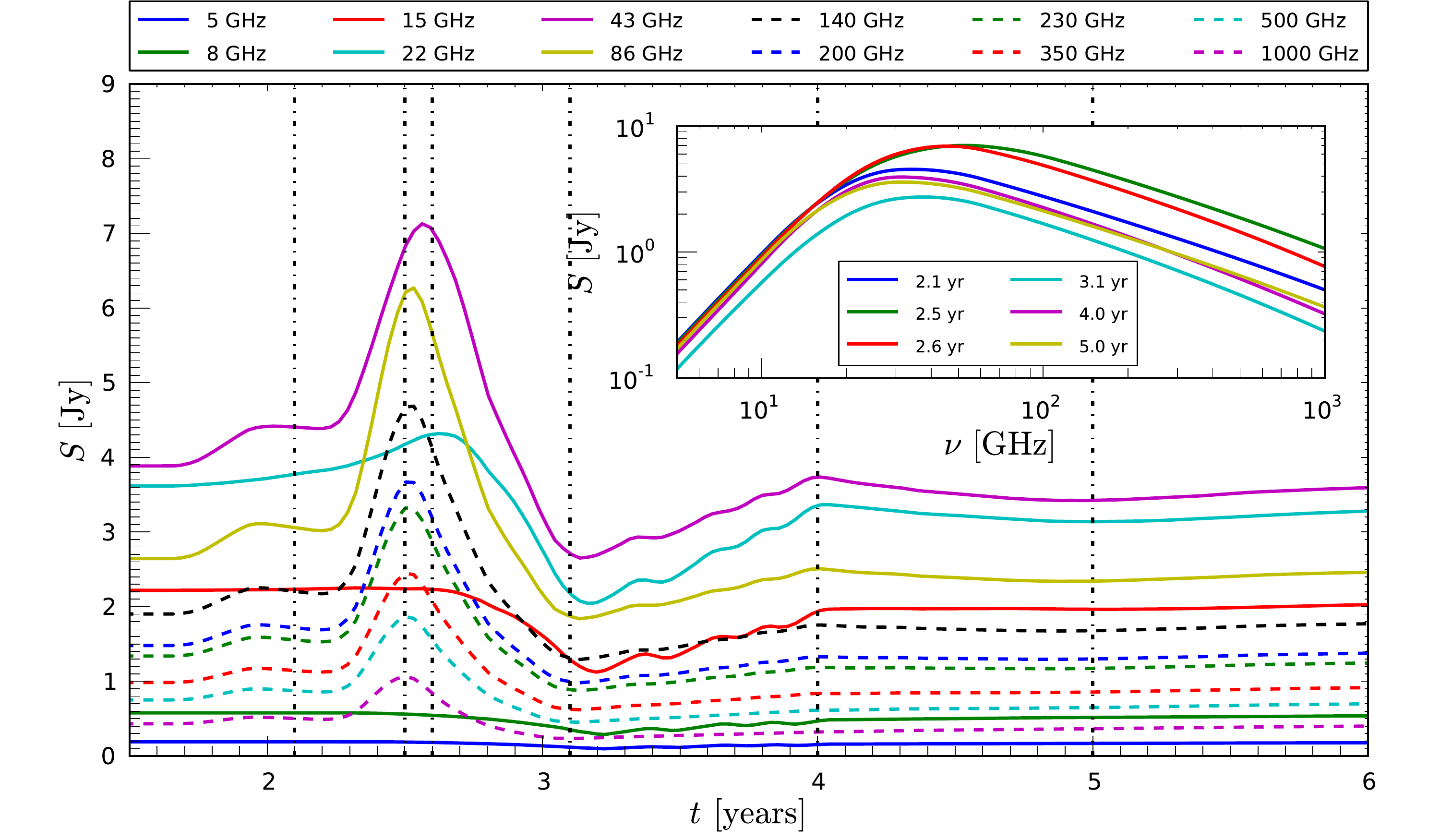} 
\caption{Same as. Fig. \ref{lcdk3} for the PM jet.}
\label{lcdk1} 
\end{figure}

Using the parameters and the technique presented above, we computed single-dish light curves by integrating the overall emission generated at a given frequency. For the detailed comparison with observations, we computed the single-dish emission for twelve frequencies that are commonly used in $mm$-$cm$ single dish observations. For the calculation of the emission we used a typical blazar viewing angle of $\vartheta=3^\circ$ \citep[it is also is the estimated viewing angle for the jet of CTA~102][see, e.g.]{2013A&A...551A..32F}. The small viewing angle leads to a Doppler boosting of the emission:
\begin{eqnarray*}
S_{\nu\,,\mathrm{steady}}&=&S_{\nu,\mathrm{init}}\delta^{2+\alpha}\quad\mathrm{(steady\,state)}\\
S_{\nu\,,\mathrm{var}}&=&S_{\nu,\mathrm{init}}\delta^{3+\alpha}\quad\mathrm{(moving\,plasma)},
\end{eqnarray*}
with Doppler factor $\delta=1/(\Gamma(1-\beta\cos\vartheta))$ and optically thin spectral index $\alpha=(1-s)/2$. Besides the Doppler boosting of the emission, the small viewing angle leads to a \emph{piling up} of the emission along the line of sight. The light rays encounter different fractions of absorption and emission depending on the physical properties of the crossed regions. In addition to the influence of the viewing angle, $\vartheta$, we have to take cosmology into account for the proper calculations of the observed emission which leads to an additional factor of $(1+z)/D_\mathrm{L}^2$ where $z$ is the redshift of the source and $D_\mathrm{L}$ its luminosity distance. In Fig.~\ref{lcdk3} we present the single-dish light curves for the OP jet and in Fig. \ref{lcdk1} for the PM jet model.
 
We list here our main results, which are discussed in detail in Sect.~\ref{disc}. The light curves corresponding to both the OP and PM jets show one major flare, located at $t=3.1\,\mathrm{yr}$ in the case of the OP jet and around $t=2.6\,\mathrm{yr}$ for the PM jet. Both flares are followed by a strong drop in the emission before a steady state is reached. This time span is around 1.5\,yr for the OP jet and roughly 1\,yr for the PM jet. In addition, the main flare of the OP jet is broader and shows a break in the decaying edge of the main flare (see Fig.~\ref{lcdk3}).

\begin{figure}[t!]
\includegraphics[width=\columnwidth]{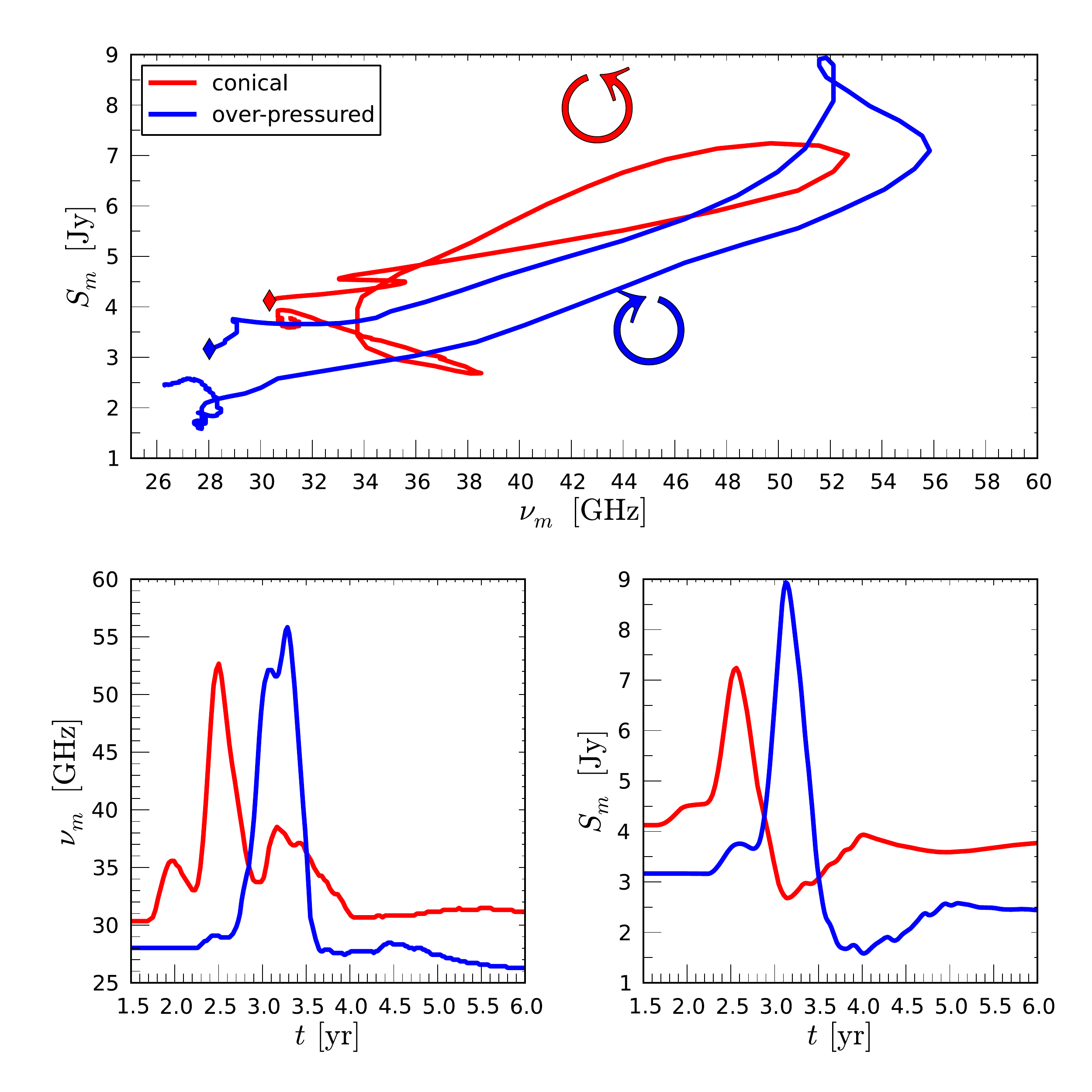} 
\caption{Evolution of the turnover values for the OP and PM jet. Top panel: Evolution of the perturbation in the turnover frequency -- turnover flux density $(\nu_m-S_m)$-plane. The diamond marker indicates the start position of the flare and the temporal evolution is indicated by the arrows. Bottom left: Temporal evolution of the turnover frequency, $\nu_m$. Bottom right: Temporal evolution of the turnover flux density, $S_m$.}
\label{vmsm} 
\end{figure}

The variation in the spectrum and the shift of the turnover point, i.e., the turnover frequency, $\nu_m$, and the turnover flux density, $S_m$, for six different times is plotted in the inlays of Fig.~\ref{lcdk3} and Fig.~\ref{lcdk1}. A more detailed evolution of the turnover over values is presented in Fig. \ref{vmsm}.

\begin{figure}[h!]
\includegraphics[width=\columnwidth]{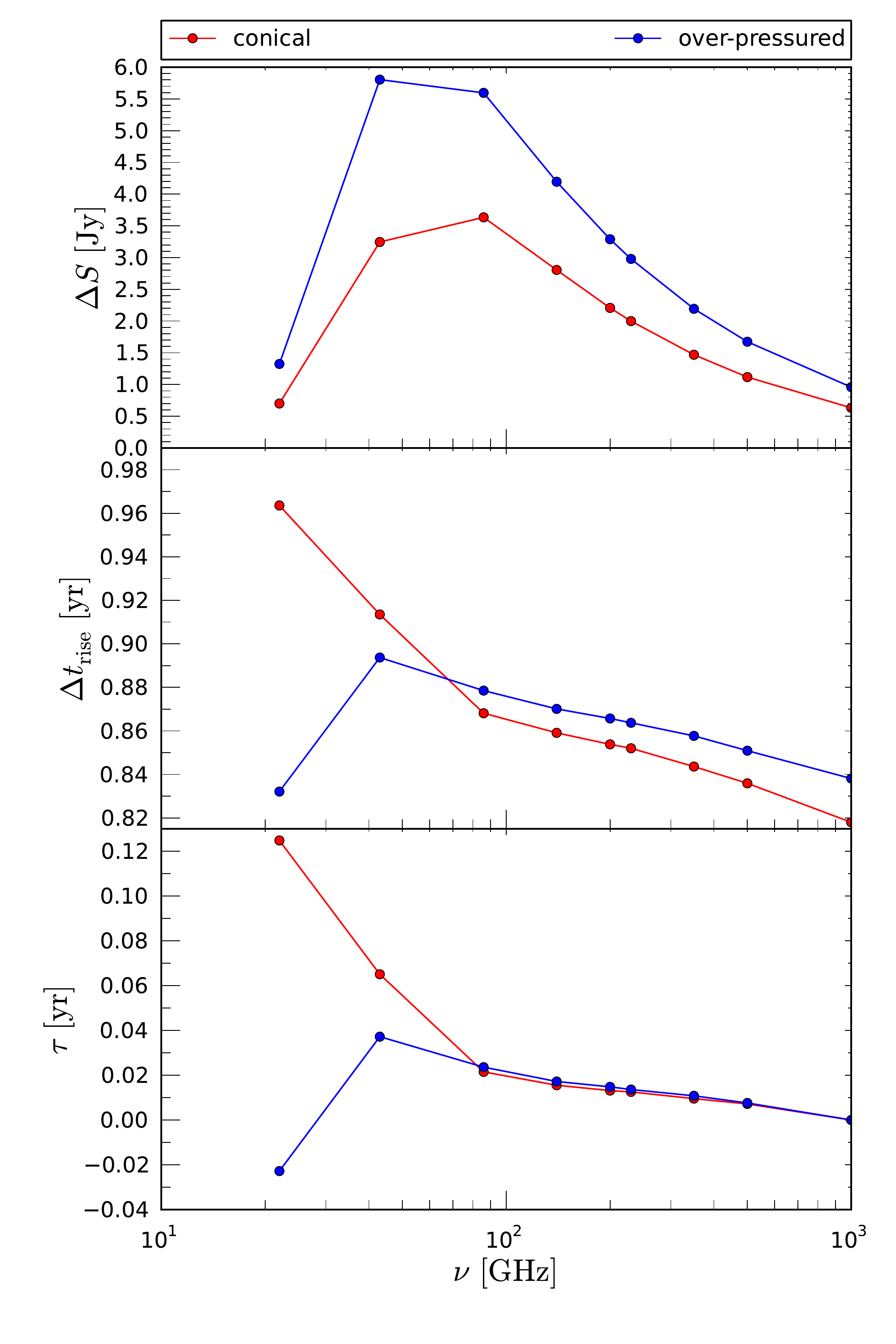} 
\caption{Single-dish light curve parameters for the PM and the OP jet. The panels show the flare amplitude, $\Delta S$, (top), the flare time scale, $\Delta t_\mathrm{rise}$, (middle) and the cross-band delays, $\tau$, with respect to the peak of 1~THz light curve (bottom).}
\label{lightcurvepara} 
\end{figure}

Figure \ref{lightcurvepara} shows single-dish light curve parameters that can help to quantify different aspects of the flare \citep[see, e.g.,][]{2014arXiv1412.7194F}. The parameters are the following: 

\begin{itemize}
\item {\bf Flare amplitude}: In order to characterise the
  frequency-dependent strength of the synthetic flares, we compute the
  light curve standard deviations $\Delta {S}$ from the ground values.

\item {\bf Flare time scale}: The time scale of a flare can be obtained from the rising 
and/or from the decaying edge of the individual light curves.
The rising time scale of a given flare is  obtained as $\Delta t=t_\mathrm{max}-t_\mathrm{min,r}$, with the time between the start of flux density increase ($t_\mathrm{min,r}$)
  and the time at the flare maximum ($t_\mathrm{max}$).

\item {\bf Cross-band delay}: In general, there is a delay between
  the flux density peaks at different frequencies. In order to
  quantify these multi-frequency {delays}, we compute the time
  difference between the flux density peak at our highest frequency
  (1000\,GHz) and the flux density peaks of the other frequencies.

\end{itemize}

\subsubsection{Synthetic radio maps}
\label{syntheticradio}
The results presented in Sect.~\ref{sec:single-dish} have been obtained by integrating the synthetic radio maps. In this section we explain the features in the radio maps themselves.

Figure~\ref{15-22_PM} shows the PM jet radio maps. The first row of panels shows the quiescent PM jet, before a perturbation is injected through the jet nozzle (left boundary). The subsequent rows show the passage of the perturbation through the jet. The perturbation is seen as an increase in the emission traveling down the jet (second row of panels), but it also leaves intermittent brightness regions behind it because of the decrease in emission in the reverse rarefaction wave (see Sect.~\ref{hydro}). As it was discussed in Sect.~\ref{sec:single-dish}, the flare is not seen at $15.4$ GHz, but it is clearly seen at $22.4$ GHz, where it peaks at approximately $2.6$ years (between the second and third rows of panels in Fig.~\ref{15-22_PM}). The flare is even more pronounced at $43.4$ GHz (third column of panels in Fig.~\ref{15-22_PM}). At the start of the flare ($t\sim 2.4$ yr, second row in the figure) the spectral index in the core increases, and as the component progresses down the jet and interacts with the previously quiescent fluid the spectrum hardens there as well (this is better observed between $22.4$ GHz and $44.4$ GHz in Fig.~\ref{15-22_PM}). In the reverse rarefaction (in the jet reference frame) wave behind the perturbation the opposite is observed: the spectrum softens, and the spectral index eventually returns to its quiescent value.

In the quiescent OP jet (top row of panels in Fig.~\ref{15-22_OP}) we observe the increase in the emission at the position of the first cross-shock ($x\sim 8.5 R_j$). The spectrum there is harder than in the rest of the jet. In fact, at $43$ GHz (third column in Fig.~\ref{15-22_OP}), most of the emission from the quiescent jet comes from the cross-shock, while at lower frequencies (left panel in Fig.~\ref{15-22_OP} the emission there is strongly absorbed. Once the perturbation enters the jet (second row of panels in Fig.~\ref{15-22_OP}), it initially behaves very similarly as in the PM jet, i.e., it is seen as an increase in the emission and in spectral index. However, once the perturbation reaches the cross-shock (fourth row of panels), the region of high emission becomes geometrically smaller, and the intensity of its radiation increases (especially at $22.4$ and $43.4$ GHz). After the perturbation passes through the cross-shock, it produces trailing components in its wake. However, in contrast to the PM case, the location of these components is in clear relation to the location of the cross-shock and not to the position of the main perturbation. Another clear difference between OP and PM jet is that the observed main component splits in two parts \citep[fourth and fifth row of panels, see e.g.,][for a thorough discussion]{2003ApJ...585L.109A,Mimica:2009de}.

\begin{landscape}
\centering
\begin{figure}
\resizebox{\hsize}{!}{\includegraphics[width=1.8\textwidth]{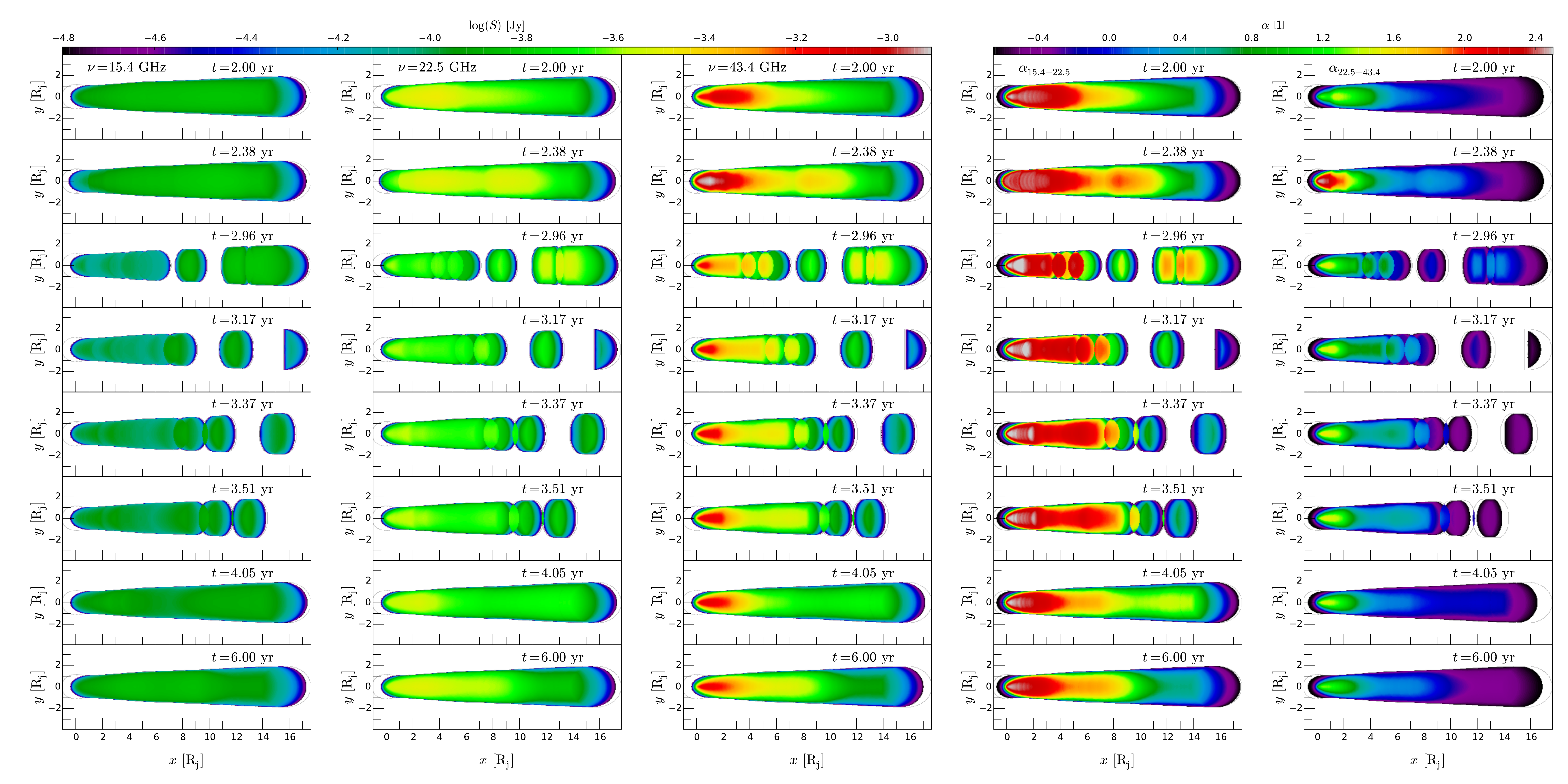}}
\caption{Synthetic observations of the PM jet at $15.4$ GHz (first column) , $22.4$ GHz (second column), and  $43.4$ GHz (third column) seen at $3^\circ$ observing angle for eight different epochs. The the forth and the fifth column right show the spectral index between the frequencies. The quiescent jet is displayed in the first row, while the subsequent rows show the passage of the perturbation and the eventual reestablishment of the quiescent jet after the perturbation exists the grid (last row).}
\label{15-22_PM} 
\end{figure}
\end{landscape}

\begin{landscape}
\centering
\begin{figure}
\resizebox{\hsize}{!}{\includegraphics[width=1.8\textwidth]{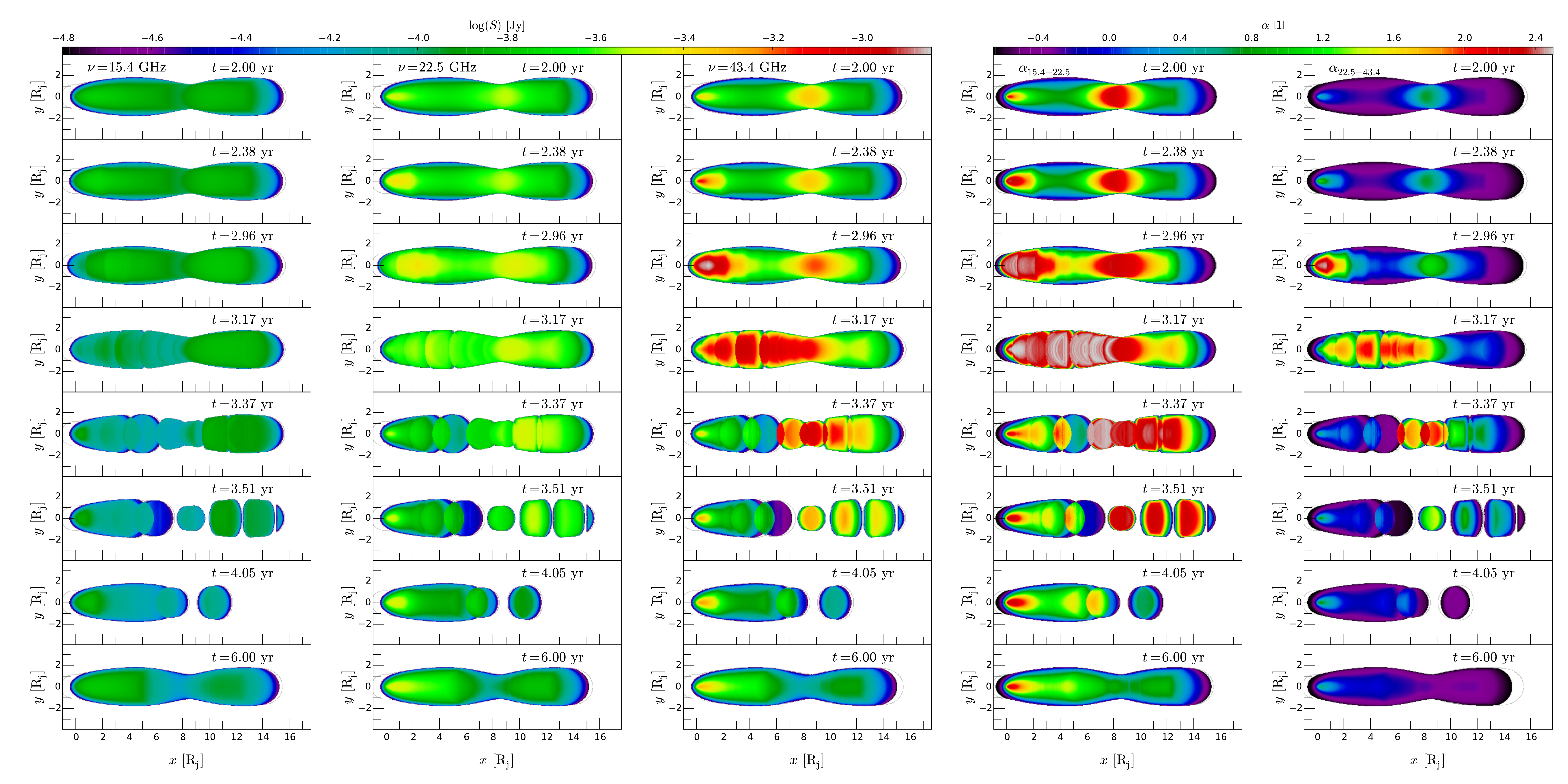}}
\caption{Same as Fig.~\ref{15-22_PM}, but for the OP jet.}
\label{15-22_OP} 
\end{figure}
\end{landscape}

\begin{figure*}[h!]
\includegraphics[width=\textwidth]{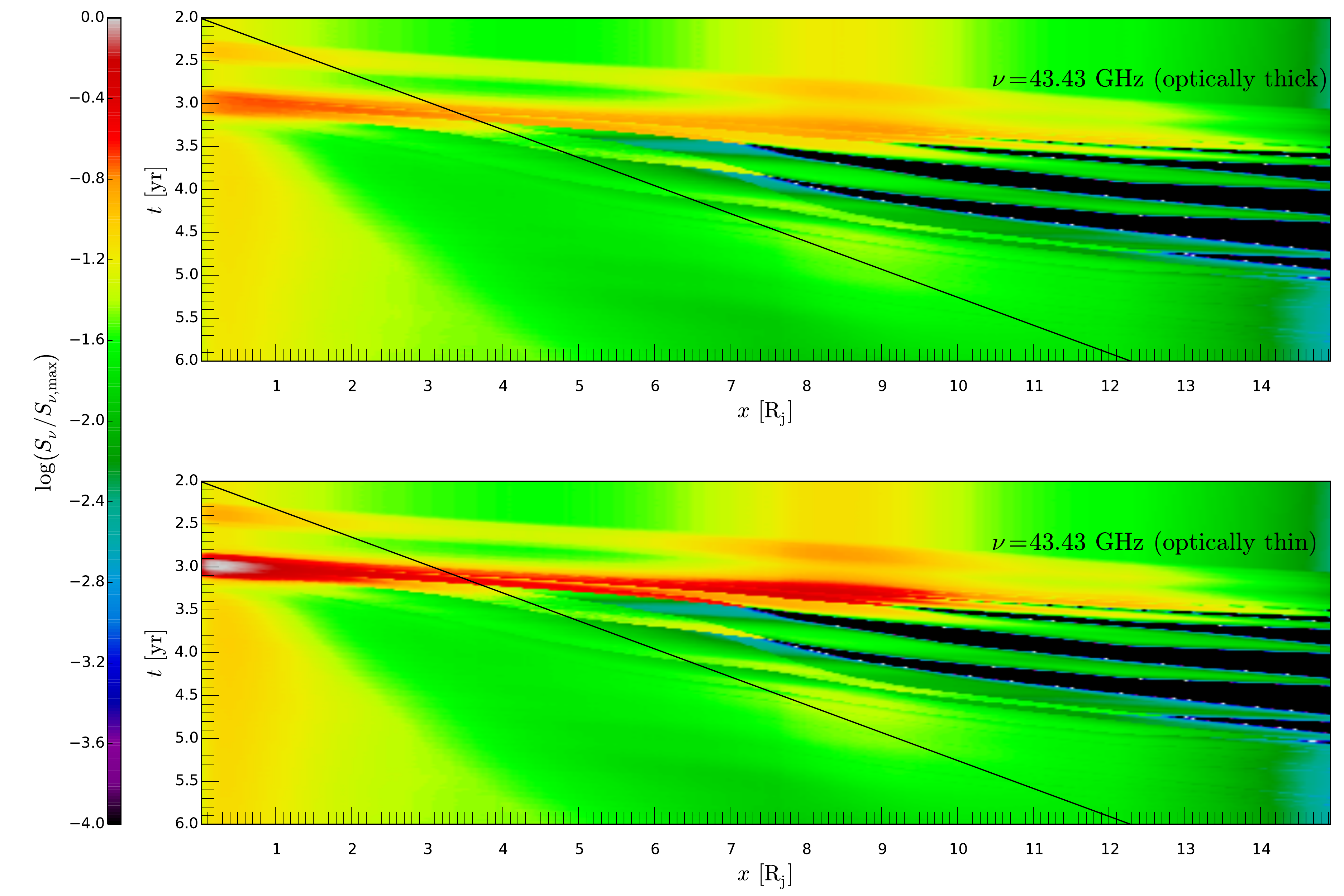} 
\caption{Time-Space plot for the variation of the axial emission at 43~GHz for the OP jet in the observer's frame at a viewing angle of $3^\circ$. The upper panel shows the emission taking opacity effects into account, whereas the bottom panel shows the optically thin emission. The solid black line indicates propagation at the speed of light, so any structure that propagates with a flatter slope is superluminal. The emission values are normalized to their maximum value (given by the optically thin map).}
\label{projection} 
\end{figure*}

 Figure~\ref{projection} shows the axial emission of the OP jet at 43~GHz with time. The upper panel shows the the emission taking opacity effects into account, whereas the bottom panel shows the optically thin emission. The optically thin emission is obviously larger than the one given when opacity is considered. In both cases, however, we observe that the time delays have an important effect on the observed emission: The emission that travels across the jet at the viewing angle collects contributions from different regions along the jet. The net effect is that the emission from the closest edge of the jet produces an initial increase of the brightness ($t\simeq 2.3-2.5\,{\rm yr}$), which is followed by a composition of emission from rarefied and shocked jet regions ($t\simeq 2.5-2.8\,{\rm yr}$) until the jet cross section from the observer's point of view. The latter corresponds to the full development of the flare in the observer's frame ($t\simeq 2.8-3.2\,{\rm yr}$).   

\subsubsection{Turnover frequency and turnover flux maps}
\label{syntheticturnover}

\begin{figure*}[t]
\includegraphics[width=\textwidth]{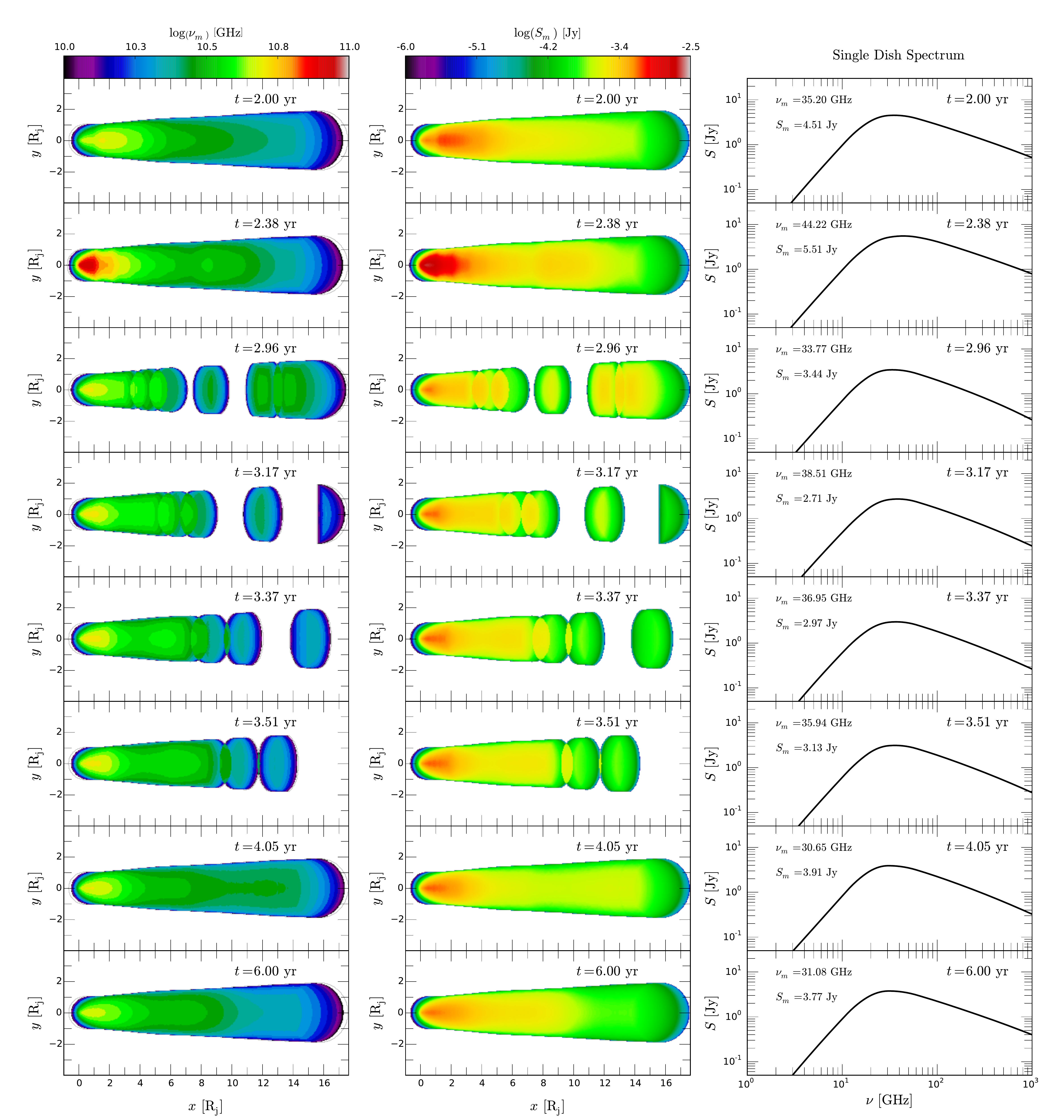} 
\caption{Turnover frequency (left column), turnover flux (middle column), and the instantaneous single-dish spectra for 8 different epochs of the PM jet observed at $3^\circ$ degrees viewing angle. The first row show the quiescent jet, while the remaining rows display the state of the jet after a perturbation has been introduced through the nozzle. In the last row the quiescent jet has almost been reestablished.}
\label{alt-PM} 
\end{figure*}

In this section we discuss the behaviour of the turnover frequency and the turnover flux density of the synthetic images. Figure~\ref{alt-PM} shows the distribution of $\nu_m$ and $S_m$ in the PM jet (left and middle panel columns, respectively), as well as the instantaneous single-dish spectrum for each of the eight epochs discussed in Sect.~\ref{syntheticradio}. The figure shows that the turnover frequency and flux steadily decrease along the quiescent jet. Once the component is introduced (second row of panels), $\nu_m$ and $S_m$ increase dramatically at the base of the jet (the fact that is also reflected in the 15\% increase in the total $\nu_m$ and $S_m$, see also red lines in Fig.~\ref{vmsm}). After the passage of the component, the quiescent jet values gradually reestablish, but the temporarily intermittent jet emission triggered by the passage of the perturbation (rows 3-6) introduces changes in the single-dish spectrum (due to missing contributions from places where the rarefaction waves decrease the emission, see Sect.~\ref{syntheticradio}).

Figure~\ref{alt-OP} shows the case of the OP jet. Here the situation is somewhat more complex, since the geometry of the jet changes along the path of the perturbation. The images at 43~GHz  (third column in Fig.~\ref{15-22_OP}) show that the biggest contribution to the emission from the OP jet comes from the cross-shock. At higher frequencies, the jet is transparent everywhere except at the very center of the cross-shock, whereas at lower frequencies it is self-absorbed ($\nu_m \simmore 30$ GHz in the cross shock and $\sim 10$ GHz elsewhere, see middle column in Fig.~\ref{alt-OP}). Once the perturbation is injected, it temporarily increases the emission at the base of the jet (top three panels of Fig.~\ref{alt-OP}, for $t\leq\,3~{\rm yr}$). A peak in $S_m$ is reached at $t=3.1,\mathrm{yr}$, when the perturbation is crossing the widest point of the jet between the nozzle and the re-collimation shock. The opacity at high frequencies is also increased by the induced increase of the flow density. Once the perturbation starts to cross the shock itself, its emitting region becomes smaller due to compression, though the opacity increases even more (see the images at $t\,=\,3.37\,{\rm yr}$ along the fifth row of the figures). The opacity in the perturbation increases, but it decreases behind the perturbation (in the larger, rarefied region). The overall effect is the net decrease of the opacity and flux. After the perturbation exits the narrow region ($t\geq\,3.5\,{\rm yr}$), its opacity starts to decrease, but the size of the emitting region increases, leading to the second peak in total opacity, followed by a decline in the peak frequency (the blue line in the left panel of Fig.~\ref{vmsm}), as the spectral peak of the source lies around 43~GHz. We note that $S_m$ does not experience the second peak since the stationary, quiescent jet establishes itself back to equilibrium already at the widest point ($x\simeq\,4\,R_j$) and the subsequent decrease of emission is not compensated by the intense emission from the post-cross-shock perturbation.

\section{Discussion}
\label{disc}
Our emission simulations do not include Compton losses, though they are deemed important in blazars. Taking this into account, we now discuss our results, and subsequently compare the synthetic single-dish emission we computed  to the radio observations of blazars flares.

\subsection{Summary of synthetic observations}
  
Figures~\ref{lcdk3} and \ref{lcdk1} show that the injection of a perturbation introduces an increase of the emission and the opacity in the simulated jets. The flux increase is larger and the duration of the flare is longer in the OP than in the PM jet. The flare in the former shows a small decrease in the slope after the main peak, followed by a steeper decrease. The emission at high frequencies is optically thin when the flare starts due to the absence of Compton scattering.

Figure~\ref{vmsm} shows that the flux density and the peak frequency increase as the shock evolves, since the amount of shocked gas increases and fills the cross-section of the jet in the observer's frame, as shown in Fig.{\ref{projection}. Therefore, the maximum flux density is observed some time after the injection of the perturbation. After this point, in the case of the PM jet, both the peak frequency and flux of the perturbed spectrum decrease as the shock expands along the jet. On the contrary, in the case of the OP jet the interaction with the standing shock changes the evolution on the $\nu_m-S_m$-plane: the flux increases in the beginning of the interaction, and instead of following a parallel track to that of the PM jet, an increase in the opacity and a decrease in the flux can be observed (see Sect.~\ref{syntheticturnover} for a detailed explanation of this feature).

 \begin{figure*}[t]
\includegraphics[width=\textwidth]{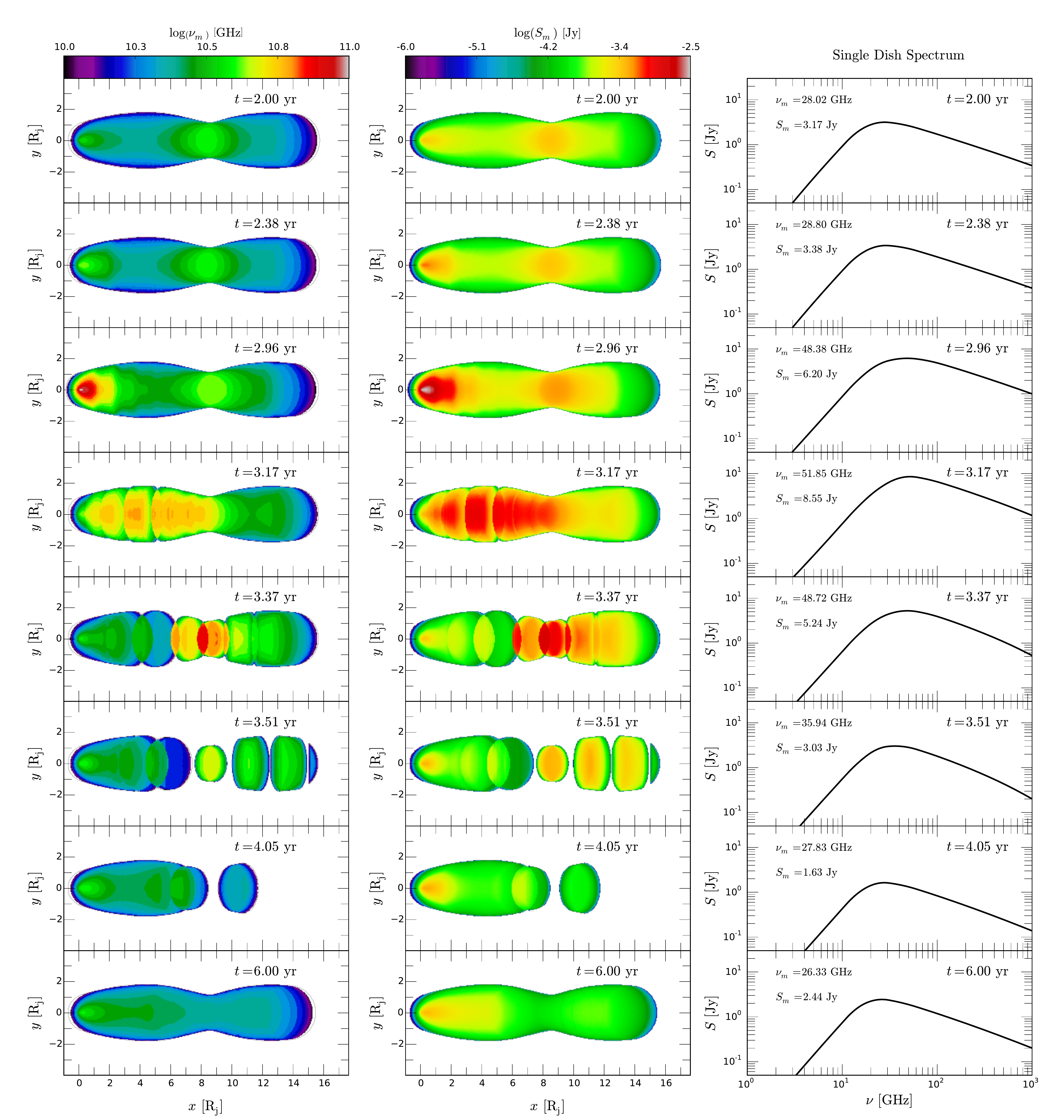} 
\caption{Same as Fig.~\ref{alt-PM}, but for the OP jet.}
\label{alt-OP} 
\end{figure*}

From Figs.~\ref{alt-PM} and \ref{alt-OP}, which show the map of the distribution of peak frequency (left column) and peak flux (right column), we can easily see that the cross-shock of the OP jet completely changes the temporal and spatial properties of the spectral distribution. It introduces a delay in the peaks with respect to the PM jet (also seen observed in Fig.~\ref{vmsm}). The outcome of the interaction between the traveling perturbation and the standing shock is the production of a stronger, delayed peak, which can be clearly distinguished from the one produced by a perturbation propagating through a PM jet.

\subsubsection{Adiabatic and radiative losses}

As in the case studied by \citet{Mimica:2009de}, the losses in the simulated regions are dominated by adiabatic losses, which explains why we do not observe a synchrotron stage in the $\nu_m-S_m$ plane. Therefore, we limit our discussion to this fact, which influences the tracks followed by the spectral peaks in that plane (Fig.~\ref{vmsm}). We note that in the quiescent OP jet the particle populations undergo compressions when they encounter the cross-shock, but the energy gained there is not sufficient for the synchrotron losses to become dominant there. Similarly, the forward shock of the perturbation is not strong enough to cause the synchrotron loss zone to become active there. The passage of the perturbation causes a reverse rarefaction wave in the jet behind it. Through this wave, the particles experience strong adiabatic losses, a feature which can nicely be seen as a decrease (or even absence) of emission in the rows 4-6 of Fig.~\ref{15-22_PM}. We note that the observed spectral index also decreases. Although the adiabatic losses cannot change the \emph{electron distribution} spectral index, the strong adiabatic cooling in the rarefaction can cause almost all the particle distribution characteristic frequencies to fall below the observational frequency (especially at $43.4$ GHz, see Fig.~\ref{15-22_PM}). Once the quiescent jet reestablishes itself, both the emission and the spectral index quickly return to the original values.
   
\subsection{Comparison with theoretical models}

   Comparing the evolution of the single-dish light-curve parameters obtained from the simulations and those expected from purely theoretical modeling \citep[Fig.~\ref{lightcurvepara}, see also][]{2014arXiv1412.7194F}, we can see that: 
 \begin{enumerate}
\item The flare amplitudes in the simulations behave in the same expected qualitative way as predicted by the theoretical modeling, showing a peak at tens of GHz;
\item The flare time-scales show a continuous decrease in the PM simulation as predicted by the theoretical modeling, but in the case of the OP jet we observe an inversion of this time-scale at 43~GHz;
\item The cross-band delays show the same qualitative behaviour for the PM and theoretical modeling, i.e., a decrease in the value of this delay from lower to higher frequencies, but an inversion in the OP jet at 43~GHz. 
\end{enumerate}

Actually, one of the main differences between both models is the fact that the peak in the OP jet occurs at 22~GHz before than at any other frequency. At frequencies $\nu \geq 90$~GHz both the PM and the OP models produce a similar qualitative evolution (the lines are basically parallel in the top and mid panels, and coincide in the bottom panel). The main quantitative differences between both simulations occur precisely at the frequencies within which the spectral peak oscillates (see Fig.~\ref{vmsm}).

   The different quantitative jump in flux increase and the delay in the peaks between both models (bottom panels in Fig.~\ref{vmsm}) show that the presence of a recollimation shock changes the spectral evolution of a perturbation. Whereas in the case of the PM model, the flux increase is purely dominated by the injection, in the OP case, the flare is delayed and shows a stronger increase in flux. Both the delay and the increase in flux can only be attached to the interaction of the perturbation with the standing shock, as discussed in the previous section. Therefore, we should expect stronger flares and a sudden increase in flux for the case of flares that interact with recollimation shocks. This signature can only be tested in those jets in which the interaction is resolved, i.e., in which the second peak can be clearly separated from that due to injection in the Compton/synchrotron stages, and has indeed been observed by \citet{2011A&A...531A..95F, 2013A&A...551A..32F,2013A&A...557A.105F,2014arXiv1412.7194F} and by \citet{2012ApJ...752...92A}. The comparison presented here between jets with and without recollimation shocks and the implications for the expected observations supports the interpretation given in those papers.

\subsection{Comparison with observations of CTA~102}
   
     In this subsection, we discuss the similarities and differences found when comparing the results shown in the single-dish light-curves (Fig.~\ref{lcdk3} and \ref{lcdk1}) with the observed light-curve from the 2006 flare in CTA~102 \citep[Fig. 1 in][]{2011A&A...531A..95F} and between the radio maps at different frequencies (Figs.~\ref{15-22_PM} and \ref{15-22_OP}) and VLBI observations \citep{2013A&A...551A..32F,2013A&A...557A.105F}. 
   
  Regarding flux variations, we must recall that the highest frequencies start to show an increase in flux before the low frequencies in CTA~102, whereas this does not happen in the simulated models by construction: We simulate a jet that does not include the most compact regions, thus the lack of the necessary opacity to reproduce this effect. Despite this difference, we find relevant similar behaviours in the flare evolution: All cases show a pre-bump in flux at intermediate frequencies (14.5 and 37~GHz in the case of CTA~102, 43 and 86~GHz in the case of the PM jet, and 22, 43 and 86 in the OP jet); the perturbation produces little or no increase of the flux at low frequencies, which is seen both in the light-curve of CTA~102 and in the simulated models, and the flux increase is maximum at tens of GHz in all cases. After the flare, we observe a dip in emission at all frequencies, which is also seen in the light-curve of CTA~102 and in the simulated models.

     It is difficult to separate the peak in flux produced by the injection of the perturbation and that produced by the interaction with the cross-shock in our simulations, as it all occurs within a short time and the filling factor of the perturbation is large. However, the delay in the peak observed in the OP jet would help to distinguish such an effect with respect to the first one, in real sources, if the standing shock producing this effect is located at a given distance interval from the radio-core. This second peak will necessarily produce a more homogeneous increase of the observed flux at frequencies ranging from $\simeq\,10$ to 100's of GHz, as opposed to the first peak, mainly visible at 100's of GHz. This global increase is observed in both the light-curve of CTA~102 and the simulations. 

    The main difference between the observations of CTA~102 and the simulation of the OP jet is that the second peak in CTA~102 does not imply an increase in the peak frequency \citep[between 2006.0-2006.3, the peak frequency remains basically constant, see Fig.~5 in][]{2011A&A...531A..95F}, but it increases in the simulation. This might be due to the relative contribution of the interaction region to the total emission and opacity as it would be derived by a single-dish observation. This contribution can be exaggerated in the case of the simulations.

   Finally, it is interesting to note that the observed spectral evolution of the simulated jets is comparable to that observed in a number of the so-classified types-1, 2, and 3 flares in \citet{2012JPhCS.372a2007A}, all claimed to correspond to the same physical mechanism, namely the injection of perturbations. The main difference between the different observed types is the role of the extended emission of the jet, the relative flux variations at different frequencies and the redshift of the source. As we show here, some of the different phenomenology observed in the sources covered by the F-GAMMA sample could be explained in terms of shock-shock interaction: delayed and strong peaks of emission, probably distinguishable from the Compton peak if the temporal sampling is enough, and variable opacity along the jet as the flare evolves.  
   
   The $3^\circ$ radio maps (Figs.~\ref{15-22_PM} and \ref{15-22_OP}) show that the knotty structures and flux variations obtained from the simulations are  comparable to VLBI radio maps of sources undergoing injection of radio components. This was already discussed by \citet{1997ApJ...482L..33G}, \citet{2001ApJ...549L.183A}, and \citet{Mimica:2009de}. The main differences between the PM and the OP jet come from the spectral index at the cross-shock, which shows an increase in the OP jet, as opposed to the steady decrease of the spectral index in the PM case (Figs.~\ref{15-22_PM} and \ref{15-22_OP}). This kind of behaviour can be observed using VLBI observations \citep{2013A&A...557A.105F}. It is worth mentioning that both models show sharp decreases of flux at the rarefied regions following the perturbations and the consequent trailing features, which can also be observed in VLBI maps following superluminal radio-components \citep{2001ApJ...549L.183A,Mimica:2009de}. 

Comparing the simulated $\nu_m$ - $S_m$ evolution (upper panel in Fig.~\ref{vmsm} with the observed values \citep[Fig.~20 of][]{2013A&A...557A.105F}, we note that both in simulated and in the observed jet the opacity increases following the peak of the emission (Sect.~\ref{syntheticturnover}, second paragraph). From our simulations we know that this happens far downstream the jet core, and is therefore free of any Compton and synchrotron losses. Thus, our neglecting of the Compton losses does not influence the results obtained for the shock interaction, and the mechanism we propose can be used to explain the observed phenomenology of CTA 102 \citep{2013A&A...557A.105F}.

\section{Conclusions}
   
In this work we studied blazar flares in a conical (pressure-matched) and in an over-pressured jet. We show that the injection of perturbations in a jet can produce a bump in emission at GHz frequencies previous to the main flare, which is produced when the perturbation fills the jet in the observer's frame. We also show that the flare spectral evolution can be completely changed in an over-pressured jet with respect to the conical case. The interaction between the injected perturbation and the standing shock produces a larger relative increase in the peak flux and introduces significant changes in the evolution of the peak frequency, as a result of the interplay between emissivity and absorption at the interaction region. Furthermore, for the over-pressured jet, we observe differences in the light-curve parameters: The cross-band delay is negative between 22~GHz and the fiducial frequency of 1~THz, and flaring time-scales are shorter at 22~GHz. The reason for this being that 22~GHz is the closest to the peak frequency of the spectral distribution before the injection of the perturbation, and is associated to the standing shock. These features can be better observed in interactions downstream of the core \citep{2013A&A...557A.105F}, but a detailed analysis of the light-curves during flares can provide hints of such shock interactions also within the core.

  Summarizing, the detailed analysis of our simulations and the non-thermal emission calculations show that interaction between a recollimation shock and traveling shock produce a typical and clear signature in both the single--dish light curves and in the VLBI observations: the flaring peaks are higher and delayed with respect to the evolution of a perturbation through a conical jet, and the cross-band delay and the flaring time-scales can show negative and shorter values, respectively, at tens of GHz as compared to higher frequencies (we recall that this is expected to be the opposite in coincal jets). These features can allow to detect such interactions for stationary components lying outside of the region in where the losses are dominated by inverse Compton scattering.

   Our results predict a number of observational signatures that could be detected in single-dish observations (with enough temporal sampling), and also using VLBI (Fromm et al. 2013a,b). Future work in this direction should include the analysis of a number of sources that are tracked within survey programs such as MOJAVE and show hints of interactions between traveling and standing components downstream of the radio-core (e.g., 0202+149 -4C+15.05-, 0415+379 -3C111-, 0528+134, 0738+313, 0829+046, 0851+202 -OJ287-, 1127-145, 1156+295 -4C+29.45-, 1219+285, 1253-055 -3C279-, 1418+546, 1823+568, 2200+420 -BL Lac-, 2201+315 -4C+31.63, 2230+114 -CTA102-, 2251+158 -3C454.3-). This should provide a further step in the characterisation of a scenario that has been claimed to be important to explain very-high energy emission in a number sources \citep[e.g.,][and references therein]{Agudo:2010jp,2012A&A...537A..70S} and facilitate the search of correlations between radio and gamma-ray flares. {A detailed analysis will also require full RMHD simulations in order to learn about the exact role of the magnetic field intensity and configuration on our results.}

\begin{acknowledgements}
M.P. is a member of the working team of projects AYA2013-40979-P and AYA2013-48226-C3-2-P, funded by MINECO. PM acknowledges the support from the European Research Council (grant CAMAP-259276), the government of Spain (MINECO grant AYA2013-40979-P), and the regional government of Valencia (grant PROMETEO-II-2014-069). 

\end{acknowledgements}

\bibliographystyle{aa} 
\bibliography{biblo_RHD.bib}
\end{document}